\documentclass[11pt,a4paper,twoside]{article}        
\makeatletter        
\usepackage{amsbsy}%
\usepackage{amsfonts}
\usepackage[centertags]{amsmath} 
\usepackage{amsopn}%
\usepackage{amssymb}
\usepackage{amstext}%
\usepackage{graphicx}%
\usepackage{array}%
\usepackage{boxedminipage}%
\usepackage{calc}%
\usepackage{cite}%
\usepackage[dvips]{color}%
\usepackage{curves}%
\usepackage{dsfont}%
\usepackage{epsfig,epic}
\usepackage{eepic}
\usepackage{fancyhdr}
\usepackage{float}%
\usepackage{latexsym}   
\usepackage{rotating}    
\usepackage{setspace}
\usepackage{subfigure}%
\usepackage{syntonly}   
\usepackage{theorem} 
\usepackage{wrapfig}
\usepackage[latin1]{inputenc}
\usepackage[T1]{fontenc}
\usepackage[english]{babel}
\newlength{\rilegatura}
\newlength{\simmetricspaceh}
\newlength{\fotogram}
\newlength{\sepfoto}
\newlength{\larggraph}
\newlength{\altgraph}
\DeclareMathOperator*{\Tr}{Tr}
\DeclareMathOperator*{\diam}{diam}

\DeclareMathOperator*{\slim}{s--lim}
\newcommand{\Lspace}[2]{L^{#1}_{\mu}\pt{#2}} 
\newcommand{\qed}{\rule[0pt]{7pt}{7pt}} 
\newcommand{\Cspace}[2]{{\cal C}^{#1}\pt{#2}}  
\def\Id{\mathds{1}}  
\def\IN{\mathds{N}}  
\def\IZ{\mathds{Z}}  
\def\IR{\mathds{R}}  
\def\IC{\mathds{C}}  
\def\pum{{\scriptstyle \frac{1}{2}}}  
\def\IT{{\mathds{T}}^2}

\def\Ac{{\cal C}^{0}\pt{\IT}}      
\def\Al{L^{\infty}_{\mu}\pt{\IT}}  
\def\om{\omega_\mu}  
\def\tn{\tau_{\cal N}}      
      
\def\nh{{\cal N}}      
\def\ZNZD{{(\IZ/ N \IZ)}^2}	
	        
\def\tripAS{\big(L^{\infty}_{\mu}\pt{\IT},\omega_\mu,\Theta\big)}

\def\tripQS{\big({\cal D}_\nh,\tn,\Theta_{\nh}\big)}

\def\coleq{\raisebox{0.0815\height}{$\colon$}\!\!\!\!=}         
							        
\def\eqcol{=\!\!\!\!\!\!\:\ \raisebox{0.0815\height}{$\colon$}}

\newcommand{\ZNZ}[1]{\IZ/#1\IZ}

\newcommand{\newatop}[2]{\genfrac{}{}{0pt}{}{#1}{#2}}   
							 
\newcommand{\pt}[1]{\left( #1 \right)}                   
\newcommand{\pq}[1]{\left[ #1 \right]}                   
\newcommand{\pg}[1]{\left\{ #1 \right\}}                 
\newcommand{\floor}[1]{\left\lfloor #1 \right\rfloor}             
                
\newcommand{\abs}[1]{\left|\left. #1 \right.\right|}              
\newcommand{\bk}[1]{\left\langle #1 \right\rangle}                
  
\newcommand{\bkkk}[3]{\left.\left\langle#1\left|#2\right|
                                          #3\right\rangle\right.} 
\newcommand{\ket}[1]{\left|\left. #1 \right.\right\rangle}        
\newcommand{\bra}[1]{\left\langle\left. #1 \right.\right|}        
\newcommand{\norm}[2]{{\left\Arrowvert #1 \right\Arrowvert}_{#2}} 
\newcommand{\enfasi}[1]{{\it #1}}          
\newcommand{\bs}[1]{\boldsymbol{#1}}

\newcommand{\ud}{\mathrm{d}}

\renewcommand{\Re}{\mathrm{Re}}
\renewcommand{\Im}{\mathrm{Im}}
\renewcommand{\geq}{\geqslant}       
\renewcommand{\leq}{\leqslant}

\newcommand{\Ir}{\mathds{Z}}

\newcommand{\idty}{\mathds{1}}

\newcommand{\coh}[2]{\ensuremath{|C_{#1}(#2) \rangle}}
\newcommand{\lcoh}[2]{\ensuremath{\langle C_{#1}(#2)|}}
\newcommand{\<}{\langle}
\renewcommand{\>}{\rangle}

\renewcommand{\c}[1]{\mathcal{#1}}

\newcommand{\co}[1]{\textsf{#1}}
{\theoremstyle{plain} 
\newtheorem{TT}{Theorem}
}
{\theoremstyle{plain} \theorembodyfont{\rmfamily}
} 
{\theoremstyle{plain} \theorembodyfont{\rmfamily}
} 
{\theoremstyle{break} \theorembodyfont{\rmfamily}

} 
{\theoremstyle{break} \theorembodyfont{\rmfamily}
} 
{\theoremstyle{break} \theorembodyfont{\rmfamily}
} 
{\theoremstyle{plain} \theorembodyfont{\rmfamily}
} 
{\theoremstyle{plain} \theorembodyfont{\rmfamily}
} 
{\theoremstyle{break} \theorembodyfont{\rmfamily}
} 
{\theoremstyle{break} \theorembodyfont{\rmfamily}
\newtheorem{PRS}{Properties}[section]} 
{\theoremstyle{break} \theorembodyfont{\rmfamily}
\newtheorem{LLL}{Lemma}[section]} 
{\theoremstyle{break} \theorembodyfont{\rmfamily}
\newtheorem{NNS}{Remarks}[section]
} 
{\theoremstyle{break} \theorembodyfont{\rmfamily}
} 
{\theoremstyle{break} \theorembodyfont{\rmfamily}
\newtheorem{DDD}{Definition}[section]
\newtheorem{DDS}[DDD]{Definitions}} 
{\theoremstyle{break} \theorembodyfont{\rmfamily}
\newtheorem{PPP}{Proposition}[section]
}

\newenvironment{Ventry}[1]%
	{\begin{list}{}{%
		\settowidth{\labelwidth}{#1}%
		\setlength{\leftmargin}{\labelwidth}}}%
	{\end{list}}%
\renewcommand\section{\@startsection{section}{1}{\z@}%
                            {-3.5ex \@plus -1ex \@minus -.2ex}%
                            {2.3ex \@plus.2ex}%
                            {\normalfont\Large\bfseries\boldmath}}
\renewcommand\subsection{\@startsection{subsection}{2}{\z@}%
                            {-3.25ex\@plus -1ex \@minus -.2ex}%
                            {1.5ex \@plus .2ex}%
                            {\normalfont\large\bfseries\boldmath}}
\renewcommand\subsubsection{\@startsection{subsubsection}{3}{\z@}%
                            {-3.25ex\@plus -1ex \@minus -.2ex}%
                            {1.5ex \@plus .2ex}%
                            {\normalfont\normalsize\bfseries\boldmath}} 
\renewcommand\paragraph{\@startsection{paragraph}{4}{\z@}%
                            {3.25ex \@plus1ex \@minus.2ex}%
                            {-1em}%
                            {\normalfont\normalsize\bfseries\boldmath}}
\renewcommand\subparagraph{\@startsection{subparagraph}{5}{\parindent}%
                             {3.25ex \@plus1ex \@minus.2ex}%
                             {-1em}%
                             {\normalfont\normalsize\bfseries\boldmath}}  
\renewcommand{\@seccntformat}[1]{\csname the#1\endcsname.\hspace{1em}}
\makeatother

\title{\bf Quantum dynamical entropies for discrete classical systems: a
  comparison}
\author{VALERIO CAPPELLINI\\
Dipartimento di Fisica Teorica\\
Universit\`a di Trieste\\
Strada Costiera 11, 34014 Trieste, Italy\\
valerio.cappellini@ts.infn.it}
\begin{document}    
\singlespacing
\maketitle
\begin{abstract}
\noindent On a family of classical dynamical systems on the
$2$--torus, we perform a discretization procedure similar to the
\co{A}nti--\co{W}ick quantization. Such a discretization is performed
by using a particular class of states, fulfilling
an appropriate dynamical localization property, typical of quantum
\co{C}oherent \co{S}tates. The same set of 
states is involved in the construction of a quantum entropy, that we
test on the discrete approximants; a correspondence with the
classical metric entropy of \co{K}olmogorov--\co{S}inai is found only
over time scales that are logarithmic in the discretization
parameter.\\[3ex] 
{\it Short Title}: Quantum dynamical entropies for discrete
classical systems\\[1ex]
Mathematics Subject Classification $2000$: 37D20; 54C70, 28D20, 81Q20, 81R30
\end{abstract}
\pagestyle{fancy}
\lhead[\fancyplain{}{\footnotesize \thepage \protect\hspace{5mm} \it
V.~Cappellini }]%
      {\fancyplain{}{}}
\rhead[\fancyplain{}{}]%
      {\fancyplain{}{\footnotesize \it Quantum dynamical entropies for discrete
classical systems: a comparison \hspace{5mm}\thepage}}
\chead{}\lfoot{}\cfoot{}\rfoot{}
\tableofcontents
\onehalfspacing
\section{Introduction}
\label{intro}
Under the term of \enfasi{classical chaos} goes a rich phenomenology of
classical dynamical systems on a compact phase space characterized by
a high sensitivity to initial conditions: if very small
initial errors exponentially amplify during the temporal evolution, the
systems is called 
\enfasi{chaotic}~\cite{Zas85:1,Gia89:1,Dev89:1,Wig90:1,Sch95:1,Cas95:1,Kat99:1}. 
Nevertheless, being the motion confined within a bounded region, 
the exponential divergence of trajectories has to be tested in a
finite domain. This leads to define the (maximal) coefficient of such
exponential amplification, which is called \enfasi{Lyapunov
exponent}, as $\displaystyle \xi\coleq \lim_{n
\to\infty}(1/n)\lim_{\delta\to0}\log\left({\delta_n}/{\delta}\right)$ ,
where we consider the initial error $\delta$ growing as $\delta_n$
under a discrete--time evolution.
When the amplification of errors is exponential, the Lyapunov
exponent $\xi$ is positive
and the system is classified as chaotic.\\
$\xi=0$ is typical of regular time--evolutions, but this also happens 
if we forbid $\delta$ to go to zero; indeed, $\delta_n\leq\Delta$
 and $\lim\frac{1}{n}$ vanishes.
This occurs for instance in the case of quantum
dynamical systems, where the uncertainly principle naturally endows
the phase--space with a $\hbar$--dependent granularity, and the
$\delta\rightarrow 0$ limit can not be achieved for finite $\hbar>0$,
but only if we perform the classical limit $\hbar\rightarrow 0$ before
the time 
one. Although this shows the non commutativity of the classical and 
the time
limits~\cite{Gia89:1,Cas95:1}, the temporal evolution of a finite
dimensional quantization 
compared with its classical 
counterpart exhibits a good agreement on a
time--scale bounded by the so called \enfasi{breaking time}
$\tau_\text{B}\pt{\hbar}$: usually, when the classical system is chaotic,
$\tau_\text{B}$ scales logarithmically in
$\hbar$~\cite{Zas85:1,Gia89:1,Cas95:1,Bou04:1,Fnj04:1,Sch04:1},
whereas for regular 
systems the scaling is
$\hbar^{-\alpha}$ for some $\alpha>0$~\cite{Zas85:1}.

A similar phenomena can be observed in discrete classical systems,
that are obtained for instance by forcing a classical system to live on
a square lattice of $N^2$ points, whose minimal spacing
$a=\frac{1}{N}$ acts as a lower bound for 
$\delta\rightarrow 0$: in this case $\frac{1}{N}$ plays in the discrete
domain the same role
that $\hbar$ plays in the quantum one and can be interpreted as a
quantization--like parameter.

By using this analogy of behaviours between quantum and discrete
 classical systems, the study of the latters
 result quite interesting and promising,
indeed we can get all benefits arising from classicality, that is the
 simplicity due to commutativity, and deeply inquire the
 chaotic property in this kinds of ``toy models''.

Since finite dimensional quantizations of classical dynamical systems
have an
algebraic formulation, this can be easily extended to \enfasi{discretization
procedures} when we restrict from the full matrix algebra of bounded
operators on a Hilbert space, typical of quantum systems, to a commutative
algebra of diagonal operators
describing a classical system~\cite{Cap04:1}.

A very useful tool of the semi--classical analysis of quantum systems
is represented by the use of \co{C}oherent \co{S}tates and a standard
quantization scheme, the \co{A}nti--\co{W}ick one~\cite{Bon03:1}, is
based on them: by mimicking this procedure we set up
a discretization involving  a class of states that we will refer to as
\co{L}attice \co{S}tates, suitably
defined on our Hilbert space.
Of course, in order to have a good quantization, the \enfasi{classical
limit}
$\hbar\rightarrow 0$
has to be tested~\cite{Wer95:1} and large part of this work has been
devoted to give and prove 
a consistent definition of a \enfasi{continuous limit} $N\rightarrow\infty$, suited for a
reasonable algebraic discretization scheme.\\
A first result in this direction is that 
the convergence of the discrete to the continuous dynamics is due to a very
special property of \co{L}attice \co{S}tates, that is known as
\enfasi{dynamical localization} property~\cite{Ben03:1}.

We apply our discretization procedure to a well known class of
classical systems~\cite{Kat99:1}, that are represented by
integer--matrix action on 
the $2$--torus; such systems can be rigorously divided into three 
families, namely \enfasi{hyperbolic}, \enfasi{parabolic} and
\enfasi{elliptic}, characterized by different chaotic properties.
As expected, differences in the behaviour of the breaking--times
$\tau_\text{B}\pt{N}$ (now of discrete/continuous
correspondence)
are found on the three different regimes.

The Lyapunov exponent is zero on systems with finite number of states
(both discrete 
and quantum) because it is an asymptotic quantity: an alternative
approach is to inquire the chaotic properties of a system during its
temporal evolution, and whether the system exhibits some kind of
\enfasi{finite--time chaos}. For classical
dynamical systems the Pesin--Ruelle Theorem~\cite{Man87:1} establish a bridge
between chaos and information, giving a relation between
the \co{K}olmogorov--\co{S}inai \enfasi{metric} entropy and the sum of
all positive Lyapunov 
exponent. Moreover, although the metric entropy is defined as a 
(partial) entropy production on the long run~\cite{Ale81:1,Kat99:1}, such a partial entropy
can be observed and analyzed even during the
temporal evolution, that is on finite times.

With the aim of using entropy to detect chaos, several \enfasi{quantum
dynamical entropies} have been introduced. In a recent
work~\cite{Ben03:1}, two of them, called \co{CNT} (Connes,
Narnhofer and Thirring)~\cite{Con87:1} and \co{ALF} (Alicky, Lindblad and
Fannes)~\cite{Ali94:1} 
are showed to converge to the \co{KS} invariant (but only in a joint
time and classical limit)
when applied to the
\co{A}nti--\co{W}ick quantization of the hyperbolic family of the
classical dynamical systems 
mentioned above. Only the hypothesis of dynamical localization
for \co{C}oherent \co{S}tates was used in obtaining that result.
Instead of extending such a
result to our discretization scheme, we 
directly study another quantum
dynamical entropy, constructed by means of \co{C}oherent \co{S}tates
and so called \co{CS}--quantum entropy~\cite{Slo94:1}.

What we show is that the \co{CS}--entropy
production of a discrete classical system does converge to the
\co{KS}--entropy production of the continuous limit,
but only over time scales logarithmic in the quantization--like
parameter $\frac{1}{N}$. This confirms the 
numerical results obtained in~\cite{Ben04:1} for the \co{ALF}--entropy
on a similar class of discrete systems, but within the
\co{W}eyl quantization--like scheme instead of the \co{A}nti--\co{W}ick.

Finally, we divided the \co{CS}--quantum entropy in its dynamical and
measure--dependent parts, and we show how the latter does not play a
role in the (positive) entropy rate.
\section{Classical Dynamical Systems and Phase--Space discretization}%
\label{CDS}
The typical description of a Classical Dynamical System is given by means of 
a measure space ${\cal X}$, the phase--space, endowed with the Borel
$\sigma$--algebra of its measurable subsets and a normalized 
measure $\mu$, ($\mu({\cal X})=1$).
The probability that phase--points belong to measurable subsets 
$E\subseteq{\cal X}$ is given by the ``volumes'' $\mu(E)=\int_E\mu\pt{\ud \bs{x}}$;
so the measure $\mu$ defines the statistical properties of the system and
represents a possible ``state''.

Every reversible discrete time dynamics amounts to an
invertible measurable 
map $T:{\cal X}\mapsto{\cal X}$ such that $\mu\circ T=\mu$, and to its
iterates $\{T^k \mid k\in\Ir\}$: $T$--invariance of the measure $\mu$
ensure that the state defined by $\mu$ can be taken as an equilibrium
state with respect to the given dynamics.

All phase--trajectories passing through $\bs{x}\in{\cal X}$ at time $0$
can be encoded into sequences ${\pg{T^k\,\bs{x}}}_{k\in\IZ}$~\cite{Kat99:1}.

Classical dynamical systems are thus conveniently described by
measure--theoretic
triplets $({\cal X},\mu,T)$. In particular, in the present work, we shall focus upon the following choices:\\[-5.5ex]
\begin{itemize}
\item[$\cal X$:] the $2$--dimensional
 torus $\IT={\IR}^2/{\IZ}^2 \-=\left\{\bs{x}=(x_1,x_2)\in\IR^2\
 \pmod{1} \right\}$;
\item[$\mu$:] the Lebesgue measure, $\mu(\ud\bs{x})=\ud x_1\,\ud x_2$,
on $\IT$; 
\item[$T$:] the invertible measurable transformations on $\IT$
represented by a modular matrix action, as follows:
\begin{subequations}
\label{AoDC_1ccc}
\begin{align}
T\,\pt{\bs{x}}&=
\begin{pmatrix}
\phantom{-}t_{11} & \phantom{-}t_{12}\\
\phantom{-}t_{21} & \phantom{-}t_{22}
\end{pmatrix}
\begin{pmatrix}
x_1\\
x_2
\end{pmatrix}
\ \pmod{1}\ ,\quad
\begin{matrix}
t_{\imath\jmath}\in\IZ\quad,\quad\forall \pt{\imath,\jmath}\in{\pg{1,2}}^2\\
\det\pt{T}=
t_{11}t_{22}-t_{21}t_{12}
= 1
\end{matrix}
\label{AoDC_1b}\\
T^{-1}\pt{\bs{x}}& =
\begin{pmatrix}
\phantom{-}t_{22} & - t_{12}\\
- t_{21} & \phantom{-}t_{11}
\end{pmatrix}
\begin{pmatrix}
x_1\\
x_2
\end{pmatrix}
\ \pmod{1}\ \cdot%
\label{AoDC_1b_inv}
\end{align}
\end{subequations}
\end{itemize}
\begin{quote}
\begin{NNS}{}\ \\[-7ex]%
\label{Rem_21}
\begin{Ventry}{\mdseries iii.}
\item[\mdseries i.] In the following, a point $\bs{x}$ of the torus,
will correspond to an 
equivalence class of $\IR^2$ points whose coordinates differ by
integer values;
\item[\mdseries ii.] in~\eqref{AoDC_1ccc} we use brackets
to distinguish between the mere matrix action $T\cdot\bs{x}$ and the%
$\pmod{1}$ one $T\pt{\bs{x}}$; 
\item[\mdseries iii.] 
	$T=\pt{\begin{smallmatrix} 2 & 1\\ 1 & 1
	\end{smallmatrix}}$ is known as \co{A}rnold \co{C}at
\co{M}ap~\cite{Kat99:1}, and it 
	is an element of ${\text{SL}}_2\left(\IZ\right)\subset 
	{\text{GL}}_2\left(\IZ\right)\subset
	{\text{M}}_2\left(\IZ\right)$, where
	the latter is the subset of $2\times2$
	matrices with integer entries, 
	${\text{GL}}_2\left(\IZ\right)$ the subset of
	invertible matrices and ${\text{SL}}_2\left(\IZ\right)$ the subset
	of matrices with determinant one;
\item[\mdseries iv.] the dynamics generated by
	$T\in{\text{SL}}_2\left(\IZ\right)$, that is the one we are
	focusing on,
	is called \enfasi{Unimodular Group}~\cite{Kat99:1} (\co{UMG}
	for short);
\item[\mdseries v.] since $\det\pt{T}=1$, the Lebesgue measure
$\mu$ is \enfasi{invariant} for all
$T^n\in{\text{SL}}_2\left(\IZ\right)$, $n\in\IZ$. 
\end{Ventry}
\end{NNS}
\end{quote}
In order to develop an algebraic discretization procedure as
in~\cite{Ben04:2}, it proves convenient
to follow an algebraic approach and replace $(\IT,\mu,T)$ with
the algebraic triple $\tripAS$, where
\begin{description}
	\item[$\Lspace{\infty}{\IT}$] is the (Abelian) Von~Neumann *-algebra
	of (equivalence classes of) essentially bounded functions on
	$\IT$~\cite{Bra79:1,Hew69:1}, equipped with the so-called
	essential supremum  
 norm $\|\cdot\|_\infty$~\cite{Rud87:1};
	\item[$\omega_{\mu}$] is the state (expectation) on
	$\Lspace{\infty}{\IT}$, 
	defined by the reference measure $\mu$ as
	\begin{equation}
	\om:\Lspace{\infty}{\IT}\ni
	f\longmapsto\omega_\mu(f)\coleq\int_{\IT} \mu(\ud\bs{x})\
	f(x)\in \IR^{+}\ ;
\label{omegamu}
	\end{equation}
	\item[$\Theta$] is the
        automorphism
        of $\Lspace{\infty}{\IT}$ defined by $\Theta^j\pt{f}\coleq f\circ T^j$, satisfying $\omega\circ\Theta^j=\omega$.   
\end{description}
\subsection{Discretization of phase--space}
\label{dops}
From an algebraic point of view, a discretization procedure resembles
very much quantization.  
Given the classical algebraic triple $\tripAS$, the core of a 
quantization--dequantization procedure (specifically an
$\nh$--dimensional quantization) is twofold:
\begin{itemize}
\item finding a pair of *-morphisms, ${\cal J}_{\nh,\infty}$ mapping
$\Lspace{\infty}{\IT}$ into a
finite dimensional algebra ${\cal M}_\nh$ (in general a
full $N\times N$ matrix algebra) and ${\cal
J}_{\infty,\nh}$ mapping backward ${\cal M}_\nh$ into
$\Lspace{\infty}{\IT}$; 
\item providing an automorphism $\Theta_{\nh}$, the quantum dynamics,
acting on ${\cal M}_\nh$ 
such that it approximates in a suitable sense the classical one, $\Theta$, on
$\Lspace{\infty}{\IT}$ as follows
\begin{equation}
{\cal J}_{\infty,\nh}^{\phantom{j}}\circ\Theta_{\nh}^j\circ {\cal
J}_{\nh,\infty}^{\phantom{j}}\xrightarrow[N\to\infty]{}
 \Theta_{\phantom{\nh}}^j\ \cdot 
\notag%
\end{equation}
\end{itemize}
The latter requirement can be seen as a modification of the so called
Egorov's property (see~\cite{Mar99:1}).  

A similar procedure, that we will call \enfasi{discretization}, can be
obtained if we replace the full matrix  
algebra ${\cal M}_\nh$ with a finite abelian one, namely the algebra
${\cal D}_\nh$ consisting of $N^2\times N^2$ diagonal matrices.

In order to give to elements of ${\cal D}_\nh$ the meaning of
\enfasi{discrete observables}, we define a suitable Hilbert space:
to do this , we consider a discretized version of
$(\IT,\mu,T)$ which arises by forcing the continuous classical system
to live on a square lattice $L_N\subseteq\IT$ of spacing
$\frac{1}{N}$:
\begin{equation}
L_N \coleq \pg{\frac{\bs{p}}{N} \ \Big|\  \bs{p}\in {\pt{\IZ / N
\IZ}}^2}\ ,
\label{llnn}
\end{equation}
where $\pt{\IZ / N \IZ}$ denotes the residual
class$\pmod{N}$, that is $0 \leqslant p_i \leqslant N-1$.

Now we take the ${\cal N}\coleq N^2$ points of $L_N$ as labels of the
	elements ${\pg{\ket{\bs{\ell}}}}_{\bs{\ell}\in   
	{\pt{\IZ / N \IZ}}^2}$ of an orthonormal basis (o.n.b.)
	of the ${\cal N}$ dimensional  
	Hilbert space ${\cal H}_{\nh}$, and we consider
	discrete algebraic triples $\tripQS$, consisting of
\begin{itemize}
\item[${\cal D}_{\nh}$:] an $\nh \times \nh$ matrix algebra diagonal
in the orthonormal basis introduced above;
\item[$\tn$:] the uniform state (expectation) on $\cal D_{\nh}$ defined by 
\begin{equation}
\tn:{\cal D}_\nh\ni
D\longmapsto\tn(D)\coleq\frac{1}{\nh}\Tr\pt{D}\in \IR^{+}\ ;
\label{tauenne}
\end{equation}
\item[$\Theta_{\nh}$:] an automorphism of $\cal D_{\nh}$ suitably
reproducing 
$\Theta$ when $N\longrightarrow\infty$ (see Section~\ref{AdoT}).
\end{itemize}

In particular, as the \co{A}nti--\co{W}ick quantization can be
obtained by means of \co{C}oherent \co{S}tates~\cite{Bon03:1}, a
similar \co{A}nti--\co{W}ick discretization of $\tripAS$ in $\tripQS$
can be performed~\cite{Ben04:2} once that we specified what we consider as
``\co{C}oherent \co{S}tates'' on $\cal
H_\nh$, and this is the purpose of next Section.

Intuitively, a discrete description of $\big(\IT,\mu,T\big)$ becomes finer when we increase $N$, the number 
of points per linear dimension on the grid $L_N$ in~\eqref{llnn}:
this corresponds to enlarging the dimension of the Hilbert space $\cal
H_\nh$ associate to the corresponding algebraic triple $\tripQS$.
In this sense, the lattice
 spacing $a\coleq\frac{1}{N}$ of the grid $L_N$ is a natural
\enfasi{discretization parameter} playing an analogous role to the
\enfasi{quantization parameter} $\hbar$. 
\subsection{\co{L}attice \co{S}tates on $\cal H_\nh$}
\label{PSCS}
In analogy with the the properties of quantum
\co{C}oherent \co{S}tates,  
we shall look for analogous states on the torus, that we shall call
\co{L}attice \co{S}tates~\cite{Ben04:2}. For the benefits of the
reader, we list below the set of properties which make quantum
\co{C}oherent \co{S}tates such a useful tool in semiclassical
analysis.\\[-4.5ex]   
\begin{quote}
\begin{PRS}[of \co{Q}uantum \co{C}oherent \co{S}tates]{}\ \\[-2.5ex]   
\label{coh}
 A family $\{\vert C_\nh(\bs{x})\rangle \mid \bs{x}\in\IT\}\in {\cal H}_\nh$ of vectors, 
 indexed by points
 $\bs{x}\in\IT$, constitutes a set of \co{C}oherent \co{S}tates on the
 torus if it satisfies the following requirements:
 \begin{enumerate}
 \item
  \co{Measurability}: $\bs{x} \mapsto \vert C_\nh(\bs{x})\rangle$ is
 measurable on 
 $\IT$;\\[-1ex]
 \item
  \co{Normalization}: $\|C_\nh(\bs{x})\|^2 = 1$, $\bs{x}\in\IT$;\\[-1ex]
 \item
  \co{Completeness}: $\displaystyle \nh \int_{\IT}\mu(\ud\bs{x})\, \coh{\nh}{\bs{x}}
  \lcoh{\nh}{\bs{x}} = \idty$;\\[-1ex]
 \item
  \co{Localization}: given $\varepsilon>0$ and $d_0>0$, there exists 
  $N_0(\varepsilon,d_0)$ such that for $N\ge N_0(\varepsilon,d_0)$ and
 $d_{\IT}(\bs{x},\bs{y})\ge d_0$ one has $\nh \;|\< C_\nh(\bs{x}), C_\nh(\bs{y}) \>|^2 \le \varepsilon$.
 \end{enumerate}
\end{PRS}
\end{quote}
\noindent
The symbol $d_{\IT}(\bs{x},\bs{y})$ used in the \co{localization} property 
stands for the length of the shorter segment
connecting the two points $\bs{x},\bs{y}\in \IT$, namely
we shall denote by
\begin{equation}
\label{dont}
d_{\IT}\pt{\bs{x},\bs{y}}  \coleq
\min_{\bs{n}\in{\IZ}^2} \norm{\bs{x}-\bs{y}+\bs{n}}{{\IR}^2}
\end{equation}
the distance on ${\IT}$.
\begin{quote}
\begin{NNS}[Topology of the \co{UMG} on the torus]{}\ \\[-7.5ex]
\label{distnorm}
\begin{Ventry}{\mdseries ii.}
\item[\mdseries i.] 
Notice that
$d_{\IT}\pt{\bs{a},\bs{b}}
=\norm{\bs{a}-\bs{b}}{{\IR}^2} 
\quad 
\text{if} \quad \norm{\bs{a}-\bs{b}}{{\IR}^2}\leq\frac{1}{2}$

\item[\mdseries ii.] All the automorphisms $T\in{\text{SL}}_2\left(\IZ\right)$
defined in~\eqref{AoDC_1ccc} act continuously on the torus, when the
topology is given by the distance~\eqref{dont}.
\end{Ventry}
\end{NNS}
\end{quote}
Resorting to the decomposition $\IT\ni\bs{x} =\pt{\frac{\floor{N
x_1}}{N},\frac{\floor{N x_2}}{N}} + 
\pt{\frac{\bk{N x_1}}{N},\frac{\bk{N
x_2}}{N}}
\eqcol
\frac{\floor{N \bs{x}}}{N} + \frac{\bk{N
\bs{x}}}{N}$,
where $\floor{\cdot}$ and $\bk{\cdot}$ denote the integer,
respectively fractional, part of a real number, we now make use of the
definition of the family $\vert C_\nh(\bs{x})\rangle$ of \co{L}attice
\co{S}tates given in~\cite{Ben04:2}, that consists in associating to
points of $\IT$ specific lattice points (see~\cite{Ben04:2},
Fig. $1$). \\[-3.5ex] 
\begin{quote}
\begin{DDD}[\co{L}attice \co{S}tates]{}\ \\[-2ex] 
\label{xxxnnn}
Given $\bs{x}\in\IT$, we shall denote by $\hat{\bs{x}}_N$ the element of
$\ZNZD$ given by
\begin{equation}
\hat{\bs{x}}_N=\pt{\hat{x}_{N,1},\hat{x}_{N,2}} \coleq\Big(\floor{N
x_1 + \pum}\,,\,\floor{N x_2 
+ \pum}\Big) \ ,
\label{loc_c32}
\end{equation}
and call \co{L}attice \co{S}tates on $\IT$ the vectors
$\coh{\nh}{\bs{x}}$ defined by
\begin{equation}
\IT\ni\bs{x} \mapsto  \vert C_{\nh}(\bs{x})\rangle\coleq
\ket{\hat{\bs{x}}_N}
\in {\cal H}_\nh\ \cdot
\label{CSforL1}
\end{equation}
\end{DDD}

\end{quote}
The reader can check in ~\cite{Ben04:2} that family $\pg{\vert
C_{\nh}(\bs{x})}$ satisfies Properties~\ref{coh}. In particular,
in the last proof, it is also shown that, due to our particular
choice of \co{L}attice \co{S}tates, we 
have a stronger
\co{localization} than in Property~\ref{coh}.4., namely
\begin{quote}
\begin{description} \item[$4^{\prime}.$]
  \co{Localization}: given $d_0>0$, there exists 
  $N_0(d_0)$ such that for $N\ge N_0(d_0)$ and
 $d_{\IT}(\bs{x},\bs{y})\ge d_0$ one has $\< C_\nh(\bs{x}),
  C_\nh(\bs{y}) \> = 0$\ .
\end{description}
\end{quote}
\subsection{\co{A}nti--\co{W}ick Discretization and its continuous
limit on $\IT$} 
\label{AWD}
In order to study the continuous limit and, more generally,
the quasi--continuous behaviour of $\tripQS$
when $N\to\infty$, we follow the
semi--classical technique known as \co{A}nti--\co{W}ick
quantization. Therefore, we start choosing concrete
discretization/de--discretization *-morphisms.\\[-2ex]
\begin{quote}
\begin{DDS}{}
\label{qWick}\ \\[-2ex]
 Given the family of \co{L}attice \co{S}tates $\{\vert
 C_{\nh}(\bs{x})\rangle\}\in{\cal H}_{\nh}$ of previous Section,  
 the \co{A}nti-\co{W}ick--like
 discretization scheme (\co{AW}, for short) is described by a one
 parameter family of
 (completely) positive unital map ${\cal J}_{\nh,\infty}:
 \Lspace{\infty}{\IT}\to\c 
 D_{\nh}$ 
 \begin{equation*}
 {\Lspace{\infty}{\IT}\ni} f \mapsto   
 \nh \int_{{\IT}}\mu(\ud \bs{x})\, f(\bs{x})\,
  \coh{\nh}{\bs{x}} \lcoh{\nh}{\bs{x}}=:{\cal J}_{\nh,\infty}(f)\in \cal D_{\nh}\quad .
 \end{equation*}
 The corresponding de--discretization operation is described by 
 the (completely) positive unital map ${\cal J}_{\infty,\nh}: {\cal
 D}_{\nh} 
 \to\Lspace{\infty}{\IT}$ 
 \begin{equation*}
  {\cal D}_{\nh} \ni X \mapsto 
  \<C_{\nh}(\bs{x}), X\,C_{\nh}(\bs{x})\>=:{\cal J}_{\infty,\nh}(X)(\bs{x})
 \in\Lspace{\infty}{\IT}
  \quad .
 \end{equation*}
\end{DDS}
\end{quote}
Both maps are identity preserving (unital) because
of the conditions satisfied 
by the family of \co{L}attice \co{S}tates and completely positive too,
since both $\Lspace{\infty}{\IT}$ and $\cal D_{\nh}$ are commutative
algebras. The reader can found in~\cite{Ben04:2} and~\cite{Ben03:1} a
list of simple properties of these maps, that 
incorporate minimal requests for rigorously defining the sense in
which the discrete dynamical systems 
$\tripQS$ tends to $\tripAS$, when $\frac{1}{N} \to 0$. 
\section{Discretization of the Dynamics}
\label{QOD}
\subsection{General properties of matrix actions on the plane}
\label{AATT}
The next natural step in our discretization procedure will be the
definition of a suitable discrete dynamics $\Theta_{\nh}$ on the
abelian algebra ${\cal D}_{\nh}$  
of Section~\ref{dops}. Before doing this we shall focus on some basic
properties of the (integer) matrix action on the plane, that are
\begin{equation}
\IR^2\ni\bs{x}\longmapsto T\,\bs{x}=
\begin{pmatrix}
t_{11} & t_{12}\\
t_{21} & t_{22}
\end{pmatrix}
\begin{pmatrix}
x_1\\
x_2
\end{pmatrix}
\in \IR^2,\quad
\begin{matrix}
t_{\imath\jmath}\in\IZ\quad,\quad\forall \pt{\imath,\jmath}\in{\pg{1,2}}^2\\
\det\pt{T}=
t_{11}t_{22}-t_{21}t_{12}
= 1
\end{matrix}
\notag%
\end{equation}
Note that in this Section we begin by considering integer matrices $T$,
with determinant one,   
mapping the plane onto
itself; in Section~\ref{AdoT} we will go back to actions on the torus $\IT$, as in~\eqref{AoDC_1b}.\\[-3.5ex]
\begin{quote}
\begin{DDS}[Families of matrix actions]{}\ \\[-2ex]
\label{famiglie}
We exclude from now on the cases $T=\pm\Id_2$, the identity on the plane, that are trivial.  
Depending on the trace of $T$ we have three families of maps, characterized by their spectral properties; in particular, denoting with
$t\coleq\frac{\Tr\pt{T}}{2}$ the semi--trace of $T$, the eigenvalues
are given by $t\pm\sqrt{t^2-1}$ and we have:
\begin{Ventry}{4}
\item[$\bs{\abs{t}>1}$ --- {\bf Hyperbolic family:}]
	One eigenvalue of $T$, $\lambda$, is
	greater than $1$ (in modulus) and the other one is $\lambda^{-1}$.
	In this case,
	distances are stretched along the direction of the
	eigenvector $|\bs{e}_+\rangle$,
	$T|\bs{e}_+\rangle=\lambda|\bs{e}_+\rangle$,
	contracted along 
	that of $|\bs{e}_-\rangle$, $T|\bs{e}_-\rangle=
	\lambda^{-1}|\bs{e}_-\rangle$. The (positive) Lyapunov
	exponent is given by $\xi=\log\abs{\lambda}$\ .
\item[$\bs{\abs{t}=1}$ --- {\bf Parabolic family:}]
	There is only one eigenvalue, whose modulus is equal to one, which corresponds to an eigenvector $|\bs{e}_0\rangle$.
\item[$\bs{\abs{t}<1}$ --- {\bf Elliptic family:}]
	The two eigenvalues are conjugate complex
	numbers $e^{i\phi}$ and $e^{-i\phi}$, whose corresponding 
	eigenvectors $\ket{\bs{e}_+}$ and $\ket{\bs{e}_-}$ are complex
	conjugate vectors of $\IC^2$. On the (non--orthogonal) basis
	$\pg{\ket{\bs{e}_{\text{R}}},\ket{\bs{e}_{\text{I}}}}\coleq
	\pg{\Re\pt{\ket{\bs{e}_+}},\Im\pt{\ket{\bs{e}_+}}}$, $T^n$ is 
	represented by means of the rotation matrix:
\begin{equation}R^n=
\begin{pmatrix}
	 \phantom{-}\cos\pt{n\phi} & \phantom{-}\sin\pt{n\phi}\\ 
	 -\sin\pt{n\phi} & \phantom{-}\cos\pt{n\phi}
\end{pmatrix}\quad\cdot %
\label{rotation}
\end{equation}
\end{Ventry}
\end{DDS}
\end{quote}
Before exploring the properties of the three regimes given above, we list now some more\\[-3.5ex]
\begin{quote}
\begin{DDS}{}\ \\[-2ex]
\label{palle}
Let $\displaystyle B_T(0)\coleq\pg{\bs{x}\in\IR^2\ \big\vert\ \ 
\norm{\bs{x}}{\IR^2}\leq 1 }$ be the unitary ball on the plane and
\begin{align}
B_T(p)&\coleq\pg{\bs{x}\in\IR^2\ \big\vert\ \ T^{-p}\bs{x}\in B_T(0)}
\label{pallan}
\intertext{be the $p$--evolved ball ($p\in\IZ$). Then define as}
B_T^{(n)}& \coleq \bigcup_{p=-n}^n B_T(p)
\label{pallamnn}
\end{align}
the union of all evolved balls from time $-n$ up to time $n$ ($n\in\IN$) and let
$\displaystyle D_T^{(n)} \coleq \diam\pq{B_T^{(n)}}$
be its diameter, so as 
$\displaystyle {D}_T(p) \coleq \diam\pq{{B}_T(p)}$
will be the diameter of the $p$--evolved ball ($\diam\pq{E}
\coleq
\sup_{\bs{x},\bs{y}\in E}\|\;\bs{x}-\bs{y\;}\|_{\IR^2}$). Further, we denote by
$\eta$ the largest eigenvalue of the matrix $\abs{T}=\sqrt{T^\dagger T}$. 
\end{DDS}
\end{quote}
Using this notation we now list three Propositions, one for each
family, that incorporate the main properties; a sketch of their proofs
is given in Appendix~\ref{app_000}.%
\newpage
\begin{quote}
\begin{PPP}[Hyperbolic family]{}\ \\[-1ex]
\label{prop_hyp}
Let $T$ be a matrix belonging to the hyperbolic family of
Definitions~\ref{famiglie}.\\
Without loss of generality we choose $\ket{\bs{e}_+}$ and
$\ket{\bs{e}_-}$ of in such a way that the angle $\beta$ from the
former to the latter lies in $(0,\pi)$ and we fix an
orthogonal reference system $(\bs{\hat{x}},\bs{\hat{y}})$ with
$x$-axis oriented  
along the eigenvector $\ket{\bs{e}_+}$: in such a system all orbits of
the (discrete) group 
${\pg{T^k}}_{k\in\IZ}$ lie on hyperbolas 
\begin{equation}
y^2\cos\beta -xy\sin\beta = \text{Const.}\quad\cdot
\label{orbits_hyp}
\end{equation}
The angle $\beta$, whose sine is positive according to our choice of $\ket{\bs{e}_+}$ and
$\ket{\bs{e}_-}$, is related with $\eta$ of Definitions~\ref{palle}
by
\begin{equation}
\sin\beta=\frac{\lambda-\lambda^{-1}}{\eta-\eta^{-1}}
\label{eta_beta_hyp}\quad;
\end{equation}
moreover, for every $n\in\IN$, the set
$B_T^{(n)}$ is confined into the hyperbolic region
delimited  by the four branches of the two hyperbolas
\begin{equation}
2\:y^2\cos\beta -2\:xy\sin\beta  -\pt{\cos\beta\pm 1} = 0
\label{hyp_hyp}\quad\cdot
\end{equation}
For the diameters, we have
\begin{equation}
D_T^{(n)}={D}_T(n)=\frac{\lambda^{n}-\lambda^{-n}}
{2\:\sin\beta}\pg{ 1+
\sqrt{1+{\pt{\frac{2\:\sin\beta}{\lambda^{n}-\lambda^{-n}}}}^2}}   
\label{lun_hyp}
\end{equation}
or, resorting to the expression for the Lyapunov exponent $\xi$ given in Definition~\ref{famiglie}:
\begin{equation}
\sin\beta\sinh\pg{\log\pq{D_T^{(n)}}} = \sinh\pt{n\,\xi}\quad\cdot
\label{lun_hyp2}
\end{equation}
Moreover\begin{equation}
\forall \;n\in\IN\quad,\quad D_T^{(n)}\leq\frac{\lambda^n}{\sin\beta}
\quad\quad\text{and}\quad\quad D_T^{(n)}
\xrightarrow[n\longrightarrow\infty]{}   
\frac{\lambda^n}{\sin\beta}\quad \cdot%
\label{asin_hyp}
\end{equation}
\end{PPP}\ \\[-11.5ex]
\begin{PPP}[Parabolic family]{}\ \\[-1ex]
\label{prop_par}
Let $T$ be a matrix belonging to the parabolic family of
Definitions~\ref{famiglie}.\\
We fix an
orthogonal reference system $(\bs{\hat{x}},\bs{\hat{y}})$ with
$x$-axis oriented  
along the eigenvector $\ket{\bs{e}_0}$: in such a system all orbits of
the (discrete) group 
${\pg{T^k}}_{k\in\IZ}$ lie on the
\begin{equation}
\begin{cases}
\text{line}\quad & y^{\phantom 2} = \text{Const.}\quad\text{if }t=+1
\label{orbits_par}\\
\text{two lines}\quad & y^2= \text{Const.}\quad\text{if } t=-1
\end{cases}\quad\cdot
\end{equation}
For every $n\in\IN$ the set
$B_T^{(n)}$ is confined into the stripe
delimited by the two lines
\begin{equation}
y^2 = 1\quad\cdot
\label{str_par}
\end{equation}
Resorting to $\eta$ of Definitions~\ref{palle}, we introduce a
positive real parameter  
\begin{equation}
J=\frac{\eta-\eta^{-1}}{2}
\label{J_par}
\quad
\end{equation}
that is used in the expression for the diameters, that is
\begin{equation}
D_T^{(n)}={D}_T(n)=nJ+\sqrt{n^2J^2+1} 
\label{lun_par}
\end{equation}
or, equivalently,
\begin{equation}
\sinh\pg{\log\pq{D_T^{(n)}}} = nJ\quad\cdot
\label{lun_par2}
\end{equation}
Moreover
\begin{align}
\forall
\;n\in\IN\quad,\quad D_T^{(n)}&\leq 2nJ+1%
\label{asin_par1}
\\
\intertext{and}
D_T^{(n)}&
\xrightarrow[n\longrightarrow\infty]{}   
2nJ\quad \cdot%
\label{asin_par2}
\end{align}
\end{PPP}\ \\[-10.5ex]
\begin{PPP}[Elliptic family]{}\ \\[-1ex]
\label{prop_ell}
Let $T$ be a matrix belonging to the elliptic family of
Definitions~\ref{famiglie}; if the entries of this matrix
\underline{are integer}, it holds true: 
\begin{alignat}{3}
&\forall\;n\in\IN&\quad&,\quad&{D}_T(n)& \leq\eta\quad,
\label{lun_ell_a}
\\
&\forall\;n\in\IN^{+}&\quad&,\quad&D_T^{(n)}& =\eta\quad,
\label{lun_ell_b}
\end{alignat}
where $\eta$ is the one introduced in Definitions~\ref{palle}.
\end{PPP}\ \\[-10.5ex]
\end{quote}
\subsection{Algebraic description of discretized \co{UMG}}
\label{AdoT}
Our aim is now to define a suitable discrete evolution
$\Theta_{\nh}$ on ${\cal D}_\nh$ (see Section~\ref{dops} for the
definitions), such that 
the discretized triplets $\tripQS$ converge to the continuous one $\tripAS$.

We start by introducing a new family of maps 
${\pg{U_{T}^{j}}}_{j\in\IZ}$, defined on
the torus $\IT\pt{{[0,N)}}$,
given by the action determined by the matrix $T\pmod{N}$, that is 
\begin{equation}
\IT\pt{{[0,N)}} \ni\bs{x}\longmapsto 
U_T^j\pt{\bs{x}} \coleq N\,
T^j\pt{\frac{\bs{x}}{N}}\in\IT\pt{{[0,N)}}\quad,\quad j\in\IZ\ ,%
\label{Ualphaa}  
\end{equation}
where $T\pt{\cdot}$ is the map defined
in~\eqref{AoDC_1ccc}.
The $U_T^j\pt{\cdot}$  maps are extensions of the $T^j\pt{\cdot}$ maps
on the enlarged torus $\IT\pt{{[0,N)}}$; moreover, they do map
the lattice $\ZNZD$ into itself, so as the maps $T^j\pt{\cdot}$ do it
with the lattice $L_N$ of~\eqref{llnn}.

Note that the map ${\pt{\IZ / N \IZ}}^2\ni\bs{\ell}\longmapsto
U_T\pt{\bs{\ell}}\in{\pt{\IZ / N \IZ}}^2$ is a bijection.\ \\[-4ex]
\begin{quote}
\begin{DDD}{}\ \\[-7.5ex]
\label{ThetaNSa}
	\item[$\Theta_{\nh}$] will denote the map:
	\begin{equation}
	{\cal D}_\nh\ni X \longmapsto
	\Theta_{\nh}^{\phantom{t}}\pt{X}  
	\coleq\sum_{\bs{\ell} \in {(\ZNZ{N})^2}} 
	X_{U_T\pt{\bs{\ell}},U_T\pt{\bs{\ell}}}
	\ket{\bs{\ell}}\bra{\bs{\ell}}\in{\cal D}_\nh
	\ \cdot 
\nonumber%
	\end{equation}
\end{DDD}
\end{quote}
The map $\Theta_{\nh}^{\phantom{t}}$ is a
*-automorphism of $\c D_\nh$; indeed
\begin{align}
\Theta_{\nh}^{\phantom{t}}\pt{X} 
& = \sum_{U_T^{-1}\pt{\bs{s}} \in {(\ZNZ{N})^2}} 
X_{\bs{s},\bs{s}}
\ket{U_T^{-1}\pt{\bs{s}}}\bra{U_T^{-1}\pt{\bs{s}}}=%
\notag\\
& = W_{T,N}^{\phantom{*}}\pt{\sum_{\newatop{\text{all
equiv.}}{\text{classes}}}  
X_{\bs{s},\bs{s}}
\ket{\bs{s}}\bra{\bs{s}}}W_{T,N}^{*}= 
\nonumber%
\\
& = W_{T,N}^{\phantom{*}}
\;X\;\;
W_{T,N}^{*}\ ,
\notag%
\end{align}
where the operators $W_{T,N}$, defined by linearly extending
the maps
\begin{equation}
{\cal H}_\nh\ni\big|\bs{\ell}\big\rangle\longmapsto
W_{T,N}^{\phantom{*}}\big|\bs{\ell}\big\rangle\coleq\ket{
U_T^{-1}\pt{\bs{\ell}}}\in{\cal H}_\nh
\label{aggiunta}
\end{equation} 
to $\c H_\nh$,
are unitary: $\displaystyle W_{T,N}^*\big|\bs{\ell}\big\rangle\coleq\ket{
U_T\pt{\bs{\ell}}}$.
For the same reason $\tn$ is a
$\Theta_{\nh}^{\phantom{t}}$\mbox{--invariant state.}
\section{Continuous limit of the dynamics}
\label{CCLD}
One of the main issues in the semi-classical analysis is to compare
if and how the quantum and classical time evolutions mimic each other
when the quantization parameter goes to zero.

In this paper we are instead considering the possible agreement between
the dynamics of continuous classical systems and that of a class of
discrete approximants.
In practice, in our case, we will study the difference 
\begin{equation}
\Theta^j - {\cal J}_{\infty,\nh}\circ\Theta_ {\nh}^j\circ
{\cal J}_{\nh,\infty}
\label{convergenza}
\end{equation}
which represents how much the discrete dynamics at timestep $j$
differs from the continuous one at the same timestep.

For quantum systems, whose classical limit is chaotic, the situation is
strikingly different from those with regular classical limit.
In the former case, classical and quantum mechanics agree, that is a
difference as in~\eqref{convergenza} is negligible,
only over times $j$ which scale logarithmically (and not as a power
law) in the quantization
parameter.  

As we shall see, such kind of scaling is not exclusively related with
non--commutativity; in fact, the quantization--like procedure
developed so 
far, exhibits a similar behaviour when $N\to\infty$ and we
recover $\tripAS$ as a continuous limit of $\tripQS$.
\subsection{Continuous limit of discretized \co{UMG}}
\label{ClfSM}
We want to show that the difference in~\eqref{convergenza} goes to
zero in a suitable topology, at least
on a certain time--scale. Such scales, commonly called
\enfasi{breaking times}, depend on the family of the considered map $T$.
In the following, we give three different scaling functions of $n$,
one per each family of matrix action, that will be compared with $\log N$
in the joint limits in $n$ and $N$ that we will construct in this Section.\ \\[-3ex]
\begin{quote}
\begin{DDD}{}\ \\[-2ex]
\label{bbtt}
We shall denote by $\Gamma_{T}\pt{n}$ the \enfasi{scaling
function} of time associate to a map $T$. In particular, on the
different families of
Definition~\ref{famiglie}, it is given by
\begin{equation}
\Gamma_{T}\pt{n} = 
\begin{cases}
\log\pt{\lambda^n}          & \text{for the {\bf hyperbolic family} of }T\\
\log n                      & \text{for the {\bf parabolic family} of }T\\
0                           & \text{for the {\bf elliptic family} of }T
\end{cases}\nonumber
\end{equation}
\end{DDD}
\end{quote}
We shall concretely show that the
difference~\eqref{convergenza} goes to zero 
with $N \to\infty$ in the strong topology over the Hilbert space
 $\Lspace{2}{\IT}$. More precisely, we have \ \\[-6.5ex]
\begin{quote}
\begin{TT}{}\ \\[0.5ex]
\label{propval}
 Let $\tripQS$ be a sequence of discretized dynamical systems as defined in
 Section~\ref{QOD}: for all $\gamma>1$,  
 \begin{equation}
\forall f\in \Al\quad,\quad  \slim_{\substack{j,N\to\infty\\
\Gamma_{T}\pt{j}<\frac{\log N}{\gamma}}} 
 \pt{\Theta^j- {\cal J}_{\infty,\nh}\circ\Theta_ {\nh}^j\circ
{\cal J}_{\nh,\infty}}\pt{f}
= 0\quad ,
\label{added2}
\end{equation}
where the limit is in the strong topology over the Hilbert space
 $\Lspace{2}{\IT}$. 
\end{TT}\ \\[-6.5ex]
\end{quote}
The previous Theorem indicates that the time limit and the
continuous limit do not commute in the parabolic and hyperbolic
cases. In particular, the difference  
between the discretized dynamics and the continuous one can be made
small by increasing $N$, while it becomes large beyond the time
scale $\Gamma_{T}\pt{j}\simeq \log N$. This phenomenon is the
same as in quantum 
chaos and points to discretization of phase space
(in the traditional semi--classical treatment of quantum systems),
rather than to non--commutativity, as the source
of the so--called \enfasi{logarithmic
breaking time} for hyperbolic systems. The constant $\gamma$ is a \enfasi{form factor}, which
reflects the fine 
structure of the dynamics: for instance, in the case of \co{Q}uantum
\co{C}at \co{M}aps\cite{Ben03:1}, $\gamma=2$.

For the elliptic case $\displaystyle \slim_{\substack{j,N\to\infty\\
\Gamma_{T}\pt{j}<\frac{\log N}{\gamma}}}=\slim_{\substack{j,N\to\infty\\
0<\frac{\log N}{\gamma}}}$ means $\displaystyle\slim_{j,N\to\infty}$;
$0<\log N$ is just a way to write that we do not consider any relation between $j$ and $N$. We adopted this, in order to have uniformity among the notations in the three different family of matrix action.

The constraint $j\le C \log \nh$ is typical of \enfasi{hyperbolic}
behaviour with Lyapunov exponent $\log\lambda$ and comes heuristically as
follows: the expansion of an 
initial small distance $\delta$ can be exponential until the distance
becomes the largest possible, namely $\delta
\lambda^{T_{\text{B}}}\simeq 1$ (on the torus). 
After discretization, the minimal distance gives $\delta=\frac{1}{N}$,
therefore one estimates $T_{\text{B}}\simeq\frac{\log N}{\log
\lambda}$, which is called \enfasi{breaking time} and sets the
time--scale over which continuous and discretized dynamics mimic each
other. 

In quantum chaos, the semi--classical analysis leads to an
estimate of $T_{\text{B}}$ exactly as above; further, 
the logarithmic dependence on $\hbar$ of
$T_{\text{B}}$ is a signature of the hyperbolic character of the
classical limit. Conversely, if the classical limit is regular (parabolic and elliptic case), then
the time scale when quantum and classical behaviours are more or less
indistinguishable goes in general as $\hbar^{-b},\ b>0$. 

The proof of Theorem~\ref{propval}
consists of several steps,
among which the most important is a property, satisfied by our choice
of \co{L}attice \co{S}tates, which we shall call
\co{Dynamical Localization}.
We give a full proof that the \co{L}attice \co{S}tates
satisfies such property, since it represents 
a natural request that should be fulfilled
by any consistent discretization/de--discretization
(quantization/de--quantization) scheme; before giving the statement of
the \co{dynamical localization} condition, let us introduce one more \\[-3.5ex]
\begin{quote}
\begin{DDD}{}\ 
\\[-2.5ex]
\label{kappaj}
We shall denote by $K_{N,n}(\bs{x},\bs{y})$ the quantity
\begin{equation}
\notag
K_{N,n}(\bs{x},\bs{y}) := 
\big\< C_\nh(\bs{x})\:,\: W_{T,N}^{n}\, C_\nh(\bs{y})\big\>=
\big\< U_T^n\pt{\hat{\bs{x}}_N}\:,\:\hat{\bs{y}}_N\big\>
\quad,
\end{equation}
where $W_{T,N}^{j}$ is the unitary operator defined
in~\eqref{aggiunta} and $\{\vert
C_\nh(\bs{x})\rangle\}$ is the set of \co{LS} of Definition~\ref{xxxnnn}.
\end{DDD}\ \\[-7.5ex]
\begin{TT}[Dynamical localization with $\bs{\{\vert
C_\nh(\bs{x})\rangle\}}$ states]{}\ \\[1ex]
\label{dynloc2}
\!\!For every $\gamma>1$ and 
$d_0>0$, there exists $N_0=N_0(\gamma,d_0)\in\IN^+$
with the following property: if \mbox{$N > N_0$} and
$\Gamma_{T}\pt{n}<\frac{\log N}{\gamma}$,
then 
\begin{equation*}
d_{\IT}\pt{T^n\pt{\bs{x}},\bs{y}} \geq d_0 \Longrightarrow 
K_{N,n}(\bs{x},\bs{y})
=
0\ ,
\end{equation*}
for all $\bs{x},\bs{y}\in\IT$, where
$K_{N,n}(\bs{x},\bs{y})$ are those of Definition~\ref{kappaj}
and the scaling function of time
$\Gamma_{T}\pt{n}$ has been introduced in Definition~\ref{bbtt}.
\end{TT}
\end{quote}
In analogy to the quantum case, \co{dynamical
localization} is what one expects from a good choice of
states suited the study of the continuous limit: in fact, it
essentially amounts to asking that \co{LS} 
remain decently localized around the continuous trajectories
while evolving with the corresponding discrete evolution. As we shall
see this is the case only on time such that $\Gamma_{T}\pt{n}<\pt{\log N}/\gamma$.
Informally, when $N\to\infty$, the quantities $K_{N,j}(\bs{x},\bs{y})$
should behave as if
$\nh|K_{N,j}(\bs{x},\bs{y})|^2\simeq\delta(T^j\pt{\bs{x}}-\bs{y})$ and this
is the content of next Proposition~\ref{deltadirac}, that will be of
use in Section~\ref{CSDE}.

This would make the discretization analogous to
the notion of \enfasi{regular
quantization} described in Section V of~\cite{Slo94:1}.
Actually, with our choice of \co{LS}, the quantity
$K_{N,j}(\bs{x},\bs{y})$ is a Kronecker delta.\ \\[-4ex]
\begin{quote}
\begin{PPP}{}\ \\[-2ex]
\label{deltadirac}
Using the same notation of Theorem~\ref{dynloc2} we have that, for any
given real number $\gamma>1$ and 
$f\in\Al$, it holds true:
\begin{equation}
 \lim_{\substack{n,N\to\infty\\
\Gamma_{T}\pt{n}<\frac{\log N}{\gamma}}} 
\norm{\nh\int_{\IT}f\pt{\bs{y}}
{\abs{K_{N,n}\pt{\;\bs{\cdot}\;,\bs{y}}}}^2\:\mu\pt{\ud\bs{y}} 
- f\pt{T^n\pt{\;\bs{\cdot}\;}}}{2}=0\quad,
\nonumber%
\end{equation}
where $\norm{\cdot}{2}$ denotes the $\Lspace{2}{\IT}$--norm.
\end{PPP}\ \\[-7.5ex]
\end{quote}
\noindent\textbf{Proof:}\\[1.5ex]
The equation of the statement can be expressed
in terms of the discretization--dediscretization operator ${\cal
 J}_{\nh,\infty}$ and ${\cal J}_{\infty,\nh}$ of
 Definition~\ref{qWick}, the discrete evolution automorphism
 $\Theta_{\nh}$ of Definition~\ref{ThetaNSa} and the continuous one
 $\Theta$ of Section~\ref{CDS}, as follows:
 \begin{equation}
\lim_{\substack{n,N\to\infty\\
\Gamma_{T}\pt{n}<\frac{\log N}{\gamma}}} 
\Big\|
 \pt{\Theta^n- {\cal J}_{\infty,\nh}\circ\Theta_ {\nh}^n\circ
{\cal J}_{\nh,\infty}}\pt{f}\Big\|_{2}=0\quad\cdot
\nonumber%
 \end{equation}
The last equation is proved in proof of Theorem~\ref{propval} (see~\eqref{added4}).\hfill$\qed$ \\[3ex]  
In order to prove Theorem~\ref{dynloc2}, we need the following 
auxiliary result.\\[-3.5ex]
\begin{quote}
\begin{PPP}{}\ \\[-2.5ex]
\label{Lemma2}
Resorting to the distance~\eqref{dont}, $\hat{\bs{x}}_N$
of Definition~\ref{xxxnnn}, $U_T$
of~\eqref{Ualphaa} and 
$\pt{\lambda,\beta,J,\eta}$ used in Propositions~\ref{prop_hyp}--\ref{prop_ell},
the following three statements hold:

\noindent For $\bs{x}\in\IT$ and $n\in{\IN}^+$
\begin{align}
\text{1)}\quad&\text{if }T \text{ is hyperbolic and }N>
\widetilde{N}_{\text{hyp}}\pt{n}\coleq \sqrt{2}
\frac{\lambda^n}{\sin\beta}\nonumber\\   
&\text{then }
d_{\IT}\pt{T^p\pt{\bs{x}},
\frac{U_T^p\pt{\hat{\bs{x}}_N}}{N}}\leq
\frac{\widetilde{N}_{\text{hyp}}\pt{n}}{2N}
\quad,\quad\forall p\leq
n\ ;
\label{ntildehyp}
\\
\text{2)}\quad&\text{if }T \text{ is parabolic and }N>
\widetilde{N}_{\text{par}}\pt{n}\coleq 
\sqrt{2}\pt{2nJ+1}\nonumber\\   
&\text{then }
d_{\IT}\pt{T^p\pt{\bs{x}},
\frac{U_T^p\pt{\hat{\bs{x}}_N}}{N}}\leq
\frac{\widetilde{N}_{\text{par}}\pt{n}}{2N}
\quad,\quad\forall p\leq
n\ ;
\label{ntildepar}
\\
\text{3)}\quad&\text{if }T \text{ is elliptic and }N>
\widetilde{N}_{\text{ell}}\coleq 
\sqrt{2}\;\eta\nonumber\\   
&\text{then }
d_{\IT}\pt{T^p\pt{\bs{x}},
\frac{U_T^p\pt{\hat{\bs{x}}_N}}{N}}\leq
\frac{\widetilde{N}_{\text{ell}}}{2N}
\quad,\quad\forall p\leq
n\ \cdot
\label{ntildeell}
\end{align}
\end{PPP}\ \\[-11.5ex]
\end{quote}
\noindent\textbf{Proof:}\\[1ex]
For every real number $t$, we have $0\leq\bk{N t
+{1}/{2}}=N t
+{1}/{2}-\floor{N t
+{1}/{2}}<1$, so that
$\abs{t-\frac{\floor{Nt+{1}/{2}}}{N}}\leq\frac{1}{2N}$, $ 
\forall\;t\in\IR$\ . From~\eqref{loc_c32} in Definition~\ref{xxxnnn},
we derive 
\begin{equation}
 d_{\IT}\pt{\bs{x}\;,\;\frac{\hat{\bs{x}}_N}{N}}\leq\frac{1}{\sqrt{2}N}
\label{nuovopt_2}
\ \ \quad,\quad\ \ 
\forall\;\bs{x}\in\IT\quad \cdot 
\end{equation}
Let us start by proving the first statement, being the other very
similar to it. Using the definition of $U_T$ given in~\eqref{Ualphaa},
we write
\begin{equation}
\norm{T^p\pt{\bs{x}}-
\frac{U_T^p\pt{\hat{\bs{x}}_N}}{N}}{\IR^2}=\norm{T^p\pt{\bs{x}}-
T^p\pt{\frac{\hat{\bs{x}}_N}{N}}}{\IR^2}=\norm{T^p\pt{\bs{x}-
\frac{\hat{\bs{x}}_N}{N}}}{\IR^2}\ ,%
\label{linearita}
\end{equation}
where in the latter equality we applied the linearity of
$T\pt{\cdot}$. As~\eqref{asin_hyp} was the maximum allowed spreading
for the unit ball $B_T(0)$ under the action of $n$ power of the matrix
$T$, now we have
\begin{equation}
\norm{T^p\pt{\bs{x}-
\frac{\hat{\bs{x}}_N}{N}}}{\IR^2}\leq\frac{\lambda^p}{\sin\beta}\norm{\bs{x}-\frac{\hat{\bs{x}}_N}{N}}{\IR^2}\leq\frac{1}{\sqrt{2}N}\frac{\lambda^n}{\sin\beta}%
\label{linearita2}
\ ,
\end{equation}
indeed $p\leq n$ and we applied~\eqref{nuovopt_2} together with Remark~\ref{distnorm}.i. In order to replace the first norm in~\eqref{linearita} with the toral distance, we apply once more the same Remark~\ref{distnorm}.i, providing that $\frac{1}{\sqrt{2}N}\frac{\lambda^n}{\sin\beta}\leq\frac{1}{2}$, that is $N\geq N_{\text{hyp}}\pt{n}$. 

The other statement~(\ref{ntildepar}--\ref{ntildeell}) are proved in
the same way, substituting in~\eqref{linearita2} the right expression
for the diameters, given for parabolic and elliptic case
from~\eqref{asin_par1},
respectively~\eqref{lun_ell_a}.\hfill$\qed$\\[1ex]   
\noindent\textbf{Proof of Theorem~\ref{dynloc2} :}\\[1ex]
Using the definition of $\{\vert
C_\nh(\bs{x})\rangle\}$ in~\eqref{CSforL1}, we easily compute
\begin{equation}
\big<C_\nh(\bs{x})\,\big\vert\, W_{T,N}^{n}\,C_\nh(\bs{y})\big> =  
\Big\langle \hat{\bs{x}}_N\;\Big\vert\;
U_T^{-n}
\pt{\hat{\bs{y}}_N}
\Big\rangle
=\delta^{(N)}_{\;
U_T^n
\pt{\hat{\bs{x}}_N}
\:,\:\hat{\bs{y}}_N
}
\ \cdot
\label{loc_c31}
\end{equation}
Using the triangular inequality, we get
\begin{multline}
d_{\IT}\pt{\frac{U_T^n\pt{\hat{\bs{x}}_N}}{N}\,,
\,\frac{\hat{\bs{y}}_N}{N}}\geq 
d_{\IT}\pt{T^n\pt{\bs{x}}\,,\,\bs{y}} -\\
-d_{\IT}\pt{T^n\pt{\bs{x}}\,,
\,\frac{U_T^n\pt{\hat{\bs{x}}_N}}{N}} - 
d_{\IT}\pt{\frac{\hat{\bs{y}}_N}{N}\,,
\,\bs{y}}\ \cdot
\label{loc_c34}
\end{multline}
Now we split the proof and we begin by focusing on the

\noindent{\bf Hyperbolic case:}

\noindent Since
$d_{\IT}\pt{T^n\pt{\bs{x}}\,,\,\bs{y}}\geq d_0$ by hypothesis,
using~\eqref{nuovopt_2} of proof of Proposition~\ref{Lemma2} 
and~\eqref{ntildehyp}, that is
\begin{equation}
N>\widetilde{N}_{\text{hyp}}\pt{n}\ \Longrightarrow\  
d_{\IT}\pt{T^n\pt{\bs{x}},  
\frac{U_T^n\pt{\hat{\bs{x}}_N}}{N}}\leq
\frac{1}{\sqrt{2}N}\frac{\lambda^n}{\sin\beta}\ ,
\label{dYYN2}
\end{equation}
we can derive from~\eqref{loc_c34} that $d_{\IT}\pt{\frac{U_T^n\pt{\hat{\bs{x}}_N}}{N}\,,
\,\frac{\hat{\bs{y}}_N}{N}}\geq 
d_0 - \frac{1}{\sqrt{2}N}\frac{\lambda^n}{\sin\beta}-
\frac{1}{\sqrt{2}N}\ \cdot$\\
The r.h.s. of the previous inequality can always be made strictly
larger than zero, 
\begin{equation}
d_{\IT}\pt{\frac{U_T^n\pt{\hat{\bs{x}}_N}}{N}\,, 
\,\frac{\hat{\bs{y}}_N}{N}}>0\ ,
\label{dYYN4}
\end{equation}
by choosing an $N$ larger than
\begin{equation}
N_{\text{\co{M}}}\pt{n} 
=\max\pg{\frac{1}{d_0\sqrt{2}}\pt{1 + \frac{\lambda^n}{\sin\beta}}\ ,\
\widetilde{N}_{\text{hyp}}\pt{n}={\sqrt{2}}\frac{\lambda^n}{\sin\beta}}\ , 
\label{loc_c347}
\end{equation}
so that the condition on the l.h.s. of~\eqref{dYYN2} is also
satisfied.
From~\eqref{loc_c31} and~\eqref{dYYN4}, we have
\begin{equation}
N>N_{\text{\co{M}}}\pt{n}\quad\Longrightarrow\quad
\big<C_\nh(\bs{x})\,\big\vert\, W_{T,N}^{n}\,C_\nh(\bs{y})\big> =
0 \ \cdot
\label{loc_c347a}
\end{equation}
Indeed, if the toral distance between two grid points
$\pt{\hat{\bs{z}}_N,\hat{\bs{w}}_N}$ is different from zero, they can
not by equal$\pmod{N}$ and so  
the periodic Kronecker delta in~\eqref{loc_c31} vanishes. 

\noindent Since the (non--decreasing) function $N_{\text{\co{M}}}\pt{n}$
in~\eqref{loc_c347} is eventually bounded by 
$\lambda^{\gamma n}$ ($\gamma$ being strictly greater than one), we define
$\overline{n}$ as the time when $\displaystyle
N_{\text{\co{M}}}\pt{\overline{n}}=\lambda^{\gamma \overline{n}}\eqcol
N_0$,
and choose $N>N_0$. Thus, if
$0<n<\overline{n}$, then
$N>N_0=N_{\text{\co{M}}}\pt{\overline{n}}>N_{\text{\co{M}}}\pt{n}$,
whereas if $\overline{n}\leq n<\frac{1}{\gamma}\frac{\log N}{\log
\lambda}$, then $N>\lambda^{\gamma n}>N_{\text{\co{M}}}\pt{n}$
and~\eqref{loc_c347a} holds for all $0< 
n<\frac{1}{\gamma}\frac{\log N}{\log 
\lambda}$, that is $\Gamma_{T}\pt{n}<\frac{\log N}{\gamma}$ as in the statement.

\noindent{\bf Parabolic case:}

\noindent Using now~\eqref{ntildepar}, that is
\begin{equation}
N>\widetilde{N}_{\text{par}}\pt{n}\ \Longrightarrow\  
d_{\IT}\pt{T^n\pt{\bs{x}},  
\frac{U_T^n\pt{\hat{\bs{x}}_N}}{N}}\leq
\frac{1}{\sqrt{2}N}\pt{2 n J + 1}\ ,
\label{dYYN12}
\end{equation}
we earn from~\eqref{loc_c34} that $d_{\IT}\pt{\frac{U_T^n\pt{\hat{\bs{x}}_N}}{N}\,,
\,\frac{\hat{\bs{y}}_N}{N}}\geq 
d_0 - \frac{1}{\sqrt{2}N}\pt{2 n J + 1}-
\frac{1}{\sqrt{2}N}\ \cdot$\\
The r.h.s. of the previous inequality can be made strictly
larger than zero, 
by choosing an $N$ larger than
\begin{equation}
N_{\text{\co{M}}}\pt{n} 
=\max\pg{\frac{\sqrt{2}}{d_0}\pt{n J + 1}\ ,\
\widetilde{N}_{\text{par}}\pt{n}={\sqrt{2}}\pt{2 n J + 1}}\ , 
\label{loc_c1347}
\end{equation}
so that the condition on the l.h.s. of~\eqref{dYYN12} is also
satisfied.
Reasoning as for the hyperbolic case, we conclude
that~\eqref{loc_c347a} still hold true in this case and we choose
$n^\gamma$ as bounding function of the (non--decreasing)
$N_{\text{\co{M}}}\pt{n}$ of~\eqref{loc_c1347}.
  
Finally, as for the hyperbolic case, we define
$\overline{n}$ as the time when $\displaystyle
N_{\text{\co{M}}}\pt{\overline{n}}={\overline{n}}^{\gamma}\eqcol
N_0$, and choose $N>N_0$. Thus, if
$0<n<\overline{n}$, then
$N>N_0=N_{\text{\co{M}}}\pt{\overline{n}}>N_{\text{\co{M}}}\pt{n}$,
whereas if $\overline{n}\leq n<N^{\frac{1}{\gamma}}$, then $N>n^\gamma>N_{\text{\co{M}}}\pt{n}$
and~\eqref{loc_c347a} holds for all $0< 
n<N^{\frac{1}{\gamma}}$, that is $\Gamma_{T}\pt{n}<\frac{\log N}{\gamma}$ as in the statement.
\newpage
\noindent{\bf Elliptic case:}

\noindent The same strategy adopted in the previous two cases, lead
now us to define a new $N_{\text{\co{M}}}$, \underline{independent} of
$n$, given by $N_{\text{\co{M}}} 
=\max\pg{\frac{1}{d_0\sqrt{2}}\pt{\eta+1}\ ,\
\widetilde{N}_{\text{ell}}\pt{n}=\eta\sqrt{2}\ }$;
thus, for $N>N_{\text{\co{M}}}$, the periodic Kronecker delta
in~\eqref{loc_c31} vanishes. 

The absence of relation between $N$ and $n$, for
$N>N_{\text{\co{M}}}$, is expressed in the relation
$\Gamma_{T}=0<\frac{\log N}{\gamma}$, always true for all
$N$.\hfill$\qed$ \\[-3ex] 

\noindent We are finally in position to conclude with\\[-3ex] 

\noindent\textbf{Proof of Theorem~\ref{propval}:}\\[-3ex] 

We will concentrate on the case of
continuous $f$, that is $f\in\Ac\pt{\subset\Lspace{2}{\IT}}$; the
extension to essentially bounded $f$ is straightforward and can be
realized by applying Lusin's 
Theorem~\cite{Hew69:1,Rud87:1,Rie55:1}, as the reader can check
in~\cite{Ben04:2}. 

\noindent Let $f\in\Ac$ and $\displaystyle
{\text{Op}}_{j,N}\pt{f}\coleq\pt{\Theta^j- {\cal 
J}_{\infty,\nh}\circ\Theta_ {\nh}^j\circ {\cal
J}_{\nh,\infty}}\pt{f}$: notice that ${\text{Op}}_{j,N}\pt{f}$ is a
multiplication operator on $\Lspace{2}{\IT}$, but also an
$\Lspace{\infty}{\IT}$ \Big(and thus also an $\Lspace{2}{\IT}$\Big)
function. According to~\eqref{added2}, we must show that
 \begin{equation}
\forall g\in \Lspace{2}{\IT}\quad
 , \quad  
 \lim_{\substack{j,N\to\infty\\
\Gamma_{T}\pt{j}<\frac{\log N}{\gamma}}} 
\norm{\;{\text{Op}}_{j,N}\pt{f}\;g\;
}{2}
= 0\quad \cdot
\notag%
 \end{equation}
Using Schwartz's inequality first with $g$ in the class of
\enfasi{simple functions} and then
using their density in $\Lspace{2}{\IT}$, we have just to show that
 \begin{equation}
\lim_{\substack{j,N\to\infty\\
\Gamma_{T}\pt{j}<\frac{\log N}{\gamma}}} 
\norm{\;{\text{Op}}_{j,N}\pt{f}\;}{2}
= 0\quad \cdot
\label{added4}
 \end{equation}
In~\cite{Ben04:2} it is shown that
\begin{align}
\norm{\;{\text{Op}}_{j,N}\pt{f}\;}{2}^2 
& = \om\pt{{\abs{f}}^2} + \tn\pq{{\cal
J}_{\nh,\infty}\pt{f}^* {\cal J}_{\nh,\infty}\pt{f}} -
2\;{\Re}\pt{I_{j,N}\pt{f}} \ ,\nonumber
\notag%
\intertext{with}
I_{j,N}\pt{f}& \coleq \tn\pq{\Big({\cal J}_{\nh,\infty}\circ\Theta^j\Big)
\pt{f}^*\pt{\Theta_ {\nh}^j\circ {\cal
J}_{\nh,\infty}}\pt{f}}\nonumber\\
& = \displaystyle  \nh \,\int_{\IT} \mu(\ud\bs{x})\, \int_{\IT}
 \mu(\ud\bs{y})\,  \overline{f(\bs{y})}\, f(T^j \pt{\bs{x}})  |\<
  C_\nh(\bs{x}), W_{T,N}^j C_\nh(\bs{y}) \>
 |^2 \ ,\nonumber
\end{align}
and that $\tn\pq{{\cal
J}_{\nh,\infty}\pt{f}^* {\cal
J}_{\nh,\infty}\pt{f}}\longrightarrow
\om\big({\abs{f}}^2\big)$ for large $N$;
so now the strategy is to prove that also $I_{j,N}\pt{f}$
goes to $\om\big({\abs{f}}^2\big)=\int_{\IT} \mu(\ud\bs{x})
|f(\bs{x})|^2$ when $j,N\to\infty$ with
$\Gamma_{T}\pt{j}<\frac{\log N}{\gamma}\ \cdot$  We want to prove that the difference
\begin{align*}
 &\left| I_{j,N}\pt{f} -
 \int_{\IT} \mu(\ud\bs{y})\, |f(\bs{y})|^2 \right| \nonumber\\
 &=  \left| \int_{\IT} \mu(\ud\bs{x})\, \int_{\IT} \mu(\ud\bs{y})\,
 \overline{f(\bs{y})}\, 
 \bigl( f(T^j \pt{\bs{x}}) - f(\bs{y}) \bigr)\, \nh|\< C_\nh(\bs{x}), W_{T,N}^{j} C_\nh(\bs{y})\>|^2
 \right| \nonumber
\intertext{is negligible for 
large $N$:
 selecting a ball $B(T^j \pt{\bs{x}},d_0)$,
one derives}
 & \le  \left| \int_{\IT} \mu(\ud\bs{x})\,
 \int_{B(T^j \pt{\bs{x}},d_0)} \mu(\ud\bs{y})\,
 \overline{f(\bs{y})} \bigl(f(T^j \pt{\bs{x}}) - f(\bs{y})\bigr) \nh|\< C_\nh(\bs{x}),
 W_{T,N}^{j} C_\nh(\bs{y})\>|^2 \right| \\
 &+ \left| \int_{\IT} \mu(\ud\bs{x})
 \int_{{\IT}\setminus B(T^j \pt{\bs{x}},d_0)} \mu(\ud\bs{y})
 \overline{f(\bs{y})}\bigl(f(T^j \pt{\bs{x}}) - f(\bs{y})\bigr) \nh|\< C_\nh(\bs{x}),
 W_{T,N}^{j} C_\nh(\bs{y})\> |^2 \right|.
\end{align*}
Applying the mean value theorem in the first double integral, we get that 
$\exists \;\bs{c} \in B(T^j \pt{\bs{x}}, d_0)$ such that
\begin{align*}
 &\left| I_{j,N}\pt{f} -
 \int_{\IT} \mu(\ud\bs{y})\, |f(\bs{y})|^2 \right| \nonumber\\
 &\le  \int_{\IT} \mu(\ud\bs{x})\,\left| 
 \overline{f(\bs{c})}\, \bigl(f(T^j \pt{\bs{x}}) -
 f(\bs{c})\bigr)\right|\,
 \int_{B(T^j \pt{\bs{x}},d_0)} \mu(\ud\bs{y})\;\nh\;|\< {\pt{W_{T,N}^{*}}}^j C_\nh(\bs{x}),  C_\nh(\bs{y})\>|^2  \\
 &\quad+ 2 \|f\|_{0}^{\:2}\int_{\IT} \mu(\ud\bs{x})
 \int_{{\IT}\setminus B(T^j \pt{\bs{x}},d_0)} \mu(\ud\bs{y})
 \;\nh\;|\< C_\nh(\bs{x}),
 W_{T,N}^{j} C_\nh(\bs{y})\> |^2 \ ,
\intertext{where we used the uniform norm
 $\norm{\,\cdot\,}{0}$, indeed $f\in\Cspace{0}{\IT}$. Finally, using \co{completeness} and
 \co{normalization} (Properties~\ref{coh}), we arrive at the upper bound}
 &\le  
\;\|f\|_{0}\sup_{\substack{\bs{z}\in\IT\\
\bs{c}\in B(\bs{z},d_0)}}
\left|
 \bigl(f(\bs{z}) -
 f(\bs{c})\bigr)\right| %
+ 2 \;\|f\|_{0}^{\:2}\quad\nh
\sup_{\substack{\bs{x}\in\IT\\
\bs{y}\not\in B(T^j \pt{\bs{x}},d_0)}}
  |\< C_\nh(\bs{x}),
 W_{T,N}^{j} C_\nh(\bs{y})\> |^2 \ \cdot
\end{align*}
By uniform continuity, the first term can be made arbitrarily
small, provided  we choose $d_0$ small enough. For
the second integral, we use Theorem~\ref{dynloc2},
which provides us with $N_0=N_0(\gamma,d_0)$ depending on the same $d_0$
, such that the second term vanishes
for all $N>N_0$ and for all $j$ such that
$\Gamma_{T}\pt{j}<\frac{\log N}{\gamma}$.\hfill$\qed$
\section{Dynamical Entropy on Discrete Systems}
\label{QDEODS}
Dealing with hyperbolic systems, one expects the instability proper to the presence of
a positive Lyapunov exponent to correspond to some degree of
unpredictability of the dynamics: classically, the metric entropy of  
\co{K}olmogorov--\co{S}inai provides the link~\cite{For92:1}.
\subsection{A classical one: \co{K}olmogorov--\co{S}inai metric entropy}
\label{KSE}
For continuous classical systems $\pt{{\c X},\mu,T}$ such as
those introduced in 
Section~\ref{CDS}, the construction of the dynamical entropy of
\co{K}olmogorov--\co{S}inai is based  
on 
subdividing $\cal X$ into measurable disjoint subsets
${\left\{E_\ell\right\}}_{\ell=1,2,\cdots, D}$ such that $\bigcup_\ell
E_\ell={\cal X}$ which form finite partitions (coarse graining s) 
${\cal E}$.

Under the action of dynamical maps $T$ in~\eqref{AoDC_1ccc},
any given partition ${\cal E}$ evolves into $T^{-j}({\cal E})$ with atoms
$\displaystyle
T^{-j}(E_\ell)=\{\bs{x}\in{\cal X}: T^j\pt{\bs{x}}\in E_\ell\}$;
one can then form finer partitions ${\cal E}_{[0,n-1]}
\coleq\bigvee_{j=0}^{n-1}T^{j}({\cal E})
$ %
whose atoms $ E_{i_0\,i_1\cdots i_{n-1}}\coleq 
 \bigcap_{j=0}^{n-1}T^{-j}E_{i_j}$
have volumes $\mu_{i_0\,i_1\cdots i_{n-1}}\coleq\mu\left(
E_{i_0\,i_1\cdots i_{n-1}}
\right)$.\ \\[-3ex]
\begin{quote}
\begin{DDS}{}\ \\[-7.5ex]
\begin{Ventry}{2)} %
\label{stringhe}
\item[1)] We shall set $\bs{i}=\pg{i_0\,i_1\cdots i_{n-1}}$ and
denote by $\Omega_D^n$ the set of $D^n$ n\_tuples with $i_j$ taking
values in $\pg{1, 2, \cdots, D}$.
\item[2)] The symbol $\bs{\hat{\imath}}$ will indicate the string
$\bs{\hat{\imath}}\coleq\pg{i_{n-1}\,i_{n-2}\cdots
i_1i_0}\in\Omega_D^n$; 
the two string $\bs{i}$ and $\bs{\hat{\imath}}$ are related by
$i_j=\hat{\imath}_{n-1-j}$, $\forall\;j\in\pg{0,\ldots,n-1}$.
\end{Ventry}
\end{DDS}
\end{quote}
\noindent
The atoms of the partitions ${\cal E}_{[0,n-1]}$ describe segments 
of trajectories 
up to time $n$ encoded by the atoms of ${\cal E}$ that are traversed at 
successive times; the volumes $\mu_{\bs{i}}=\mu\pt{E_{\bs{i}}}$
corresponds to probabilities for the system to belong to the atoms 
$E_{i_0},E_{i_1},\cdots,E_{i_{n-1}}$ at successive times $0\leq j\leq
n-1$.
The richness in diverse trajectories, that is the degree of irregularity 
of the 
motion (as seen with the accuracy of the given coarse-graining) correspond intuitively to our idea of ``complexity'' and 
can be measured by the Shannon entropy~\cite{Ale81:1} $S_\mu({\cal E}_{[0,n-1]})\coleq-\sum_{\bs{i}\in\Omega_D^n}\mu_{\bs{i}}
\log\mu_{\bs{i}}$.

On the long run, the partition ${\cal E}$ attributes to the dynamics an entropy per 
unit time--step $h_\mu(T,{\cal E})\coleq\lim_{n\to\infty}\frac{1}{n}S_\mu({\cal E}_{[0,n-1]})$.

This limit is well defined~\cite{Kat99:1} and the ``average entropy
production'' $h_\mu(T,{\cal E})$ measure how predictable the dynamics
is on the coarse grained scale provided by the finite partition ${\cal
E}$. To remove the dependence on ${\cal
E}$, the \co{KS} entropy
$h_\mu(T)$ of $\pt{{\c X},\mu,T}$
is defined
as the supremum over all finite measurable
partitions~\cite{Kat99:1,Ale81:1} $h_\mu(T)\coleq\sup_{{\cal E}}h_\mu(T,{\cal E})$.
\subsection{Dynamics and Information in the Quantum Setting}%
\label{QDE}
From an algebraic point of view, the difference between a ``quantum'' triplet
$\pt{{\c M}, \omega, \Theta}$ describing a quantum dynamical
system and classical triplets like $\tripAS$ of Section~\ref{CDS} or
$\tripQS$ of Section~\ref{dops} is that $\omega$ and $\Theta$ are now a
$\Theta$--invariant state, respectively an automorphism over a
non--commutative (C* or Von Neumann) algebra of operators $\c
M$~\cite{Cap04:1}. 
\begin{itemize}
\item
In standard quantum mechanics the algebra ${\cal M}$ is the 
von Neumann algebra $B({\cal H})$ of all bounded linear operators 
on a suitable Hilbert space ${\cal H}$.
If ${\cal H}$ has finite dimension $D$, ${\cal M}$ is the algebra of 
$D\times D$ matrices.
\item
The typical states $\omega$ are density matrices $\rho$, namely operators 
with positive eigenvalues $\rho_\ell$ such that
$\Tr(\rho)=\sum_\ell \; \rho_\ell=1$.
Given the state $\rho$, the mean value of any observable $X\in B({\cal H})$ 
is given by $\rho(X)\coleq\Tr(\rho X)$.
\item
The $\rho_\ell$ of previous point are 
interpreted as probabilities of finding the system in the
corresponding eigenstates. The uncertainty prior to the measurement is
measured by the Von~Neumann entropy of $\rho$ given by \mbox{$H\pt{\rho}\coleq -\Tr
\pt{\rho\log\rho}= - \sum_{\ell}\rho_\ell\log\rho_\ell$\ .}
\item
The usual dynamics on ${\cal M}$ is of the form $\Theta(X)=UXU^*$, where
$U$ is a unitary operator.
If one has a Hamiltonian operator that generates the continuous group 
$U_t=\exp{i\,t\,H/\hbar}$ then $U\coleq U_{t=1}$ and the time-evolution is
discretized by considering powers $U^j$.
\end{itemize}
\noindent 
The idea behind the notion of dynamical entropy is that information
can be obtained by repeatedly observing a system in the course of its
time evolution. Due to the uncertainty principle, or, in other words,
to non-commutativity, if observations are intended to gather
information about the intrinsic dynamical properties of quantum
systems, then  non-commutative extensions of the \co{KS}-entropy
ought first to decide whether quantum disturbances produced by
observations have to be taken into account or not.

Concretely, let us consider a quantum system described by a density
matrix $\rho$ acting on a Hilbert space $\c H$.  Via the wave packet
reduction postulate, generic measurement processes may reasonably
well be described by finite sets $\c Y = \{y_0, y_1,\ldots, y_{D-1}\}$
of bounded  operators $y_j\in \c B(\c H)$ such that $\sum_j y_j^* y_j
= \idty$. These sets are called \co{partitions of unity} ({\it p.u.},
for sake of shortness) and describe
the change in the state of the system caused by the corresponding
measurement process:
\begin{equation}
\label{18}
 \rho \longmapsto \Gamma^*_{\c Y}(\rho) \coleq \sum_j y_j\, \rho\, y^*_j.
\end{equation}
It looks rather natural to rely on partitions of unity to describe
the process of collecting information through repeated observations
of an evolving  quantum system~\cite{Ali94:1}. 

Our intention is now to introduce a quantum dynamical
entropy~\cite{Slo94:1}, based and constructed by means of \co{CS}, and
apply it to our families of discretized toral automorphisms. We will
show that this 
quantity does reduce to the \co{K}olmogorov--\co{S}inai invariant, but
only for time scales bounded by the logarithm of the discretization parameter $N$.

It is worth mention that the same result has been proved
in~\cite{Ben03:1} for two differents quantum dynamical entropies
(called \co{ALF}-- and \co{CNT}--entropy) applied to finite
dimensional quantum counterparts of the hyperbolic family of \co{UMG}
that we have considered within this paper.  The only hypothesis used
in~\cite{Ben03:1} to get the above mentioned result, consisted of a
\co{dynamical localization} property analogous to the one we proved
in Theorem~\ref{dynloc2}.

As a consequence, the same results of~\cite{Ben03:1}, that is the
convergence of \co{ALF}-- and \co{CNT}--entropy to the \co{KS} one,
can be obtained also in the present framework.
\subsection{\co{CS} Quantum Entropies}%
\label{CSQDE}
In order to make the description of a quantum system closer to
that of a classical one, the most useful tool consists in using
\co{CS}. The quantum measurement process itself can be depicted in
terms of \co{CS} in such a way that classical property can be
recovered in the semi--classical limit. 

Let $\pt{{\c M},\omega,\Theta}$ be a (finite dimensional) quantum
dynamical system as the 
ones introduced in Section~\ref{QDE}, with $\nh$ denoting the
dimension of its Hilbert space ${\c H}$, and $\pt{{\c X},\mu,T}$ be its
classical counterpart, the latter endowed with a classical
partition ${\c E}={\left\{E_\ell\right\}}_{\ell=1,2,\cdots, D}$ on it
(see Section~\ref{KSE}). Introduce on such a system a family of
\co{C}oherent \co{S}tates endowed with properties~\ref{coh}.

The map
\begin{equation}
{\c I}\pt{C}\pt{\rho}\coleq\nh\int_{C}\coh{\nh}{\bs{x}}\lcoh{\nh}{\bs{x}}\
\rho\ 
\coh{\nh}{\bs{x}}\lcoh{\nh}{\bs{x}}\;\mu\pt{\ud \bs{x}}\quad%
\label{inst}%
,
\end{equation}
for a measurable subset $C\subset{\c X}$ and an operator $\rho$, is called an
\enfasi{instrument}~\cite{Slo94:1}.
The map $\rho \longmapsto{\c I}\pt{C}\pt{\rho}$
describe
the change in the state $\rho$ of the system caused by a $C$--dependent
measurement process (compare with~\eqref{18}).

If we take the expectation of ${\c I}\pt{C}\pt{\rho}$, that is
$\mu^{\pt{\rho}}\pt{C}\coleq\omega\pq{{\c I}\pt{C}\pt{\rho}}$\ ,
we get the probability that a measurement on the system by the
instrument~\eqref{inst} give values in $C$, when the pre--measurement
state is $\rho$. 
If we wonder what is the probability that several measure,
taken stroboscopically at times $t_0=0\ ,\ t_1=1\ ,\ \ldots\ ,\
t_{n-1}=n-1$, give
values in $E_{i_0},E_{i_1},\ldots,E_{i_{n-1}}$, we have to compose the
instrument action~\eqref{inst} with the temporal evolution depicted in
Section~\ref{QDE}, obtaining 
\begin{align}
{\c P}_{i_0,i_1,\ldots,i_{n-1}}^{\text{\co{CS}}}&\coleq
\mu^{\pt{\rho}}_{t_0,t_1,\ldots,t_{n-1}}\pt{E_{i_0}\times E_{i_1}\times \cdots\times E_{i_{n-1}}}=
\nonumber\\
&=\omega\pq{{\c I}\pt{E_{i_{n-1}}}\circ\Theta\circ {\c I}\pt{E_{i_{n-2}}}\circ\Theta\circ\cdots\circ{\c I}\pt{E_{i_1}}\circ\Theta\circ{\c I}\pt{E_{i_0}}
\pt{\rho}}%
\label{propmis}
\end{align}
Using in~\eqref{propmis} the expression for the dynamical evolution
$\Theta(X)=U X U^*$ together with~\eqref{inst}, and replacing the
expectation $\omega$ with the trace, (see Section~\ref{QDE}),
we obtain
\begin{multline}
{\c P}_{\bs{i}}^{\text{\co{CS}}}=
{\c P}_{i_0,i_1,\ldots,i_{n-1}}^{\text{\co{CS}}}=
\nh^n\int_{E_{i_0}}\int_{E_{i_1}}\cdots\int_{E_{i_{n-1}}}
\bkkk{C_\nh\pt{\bs{x}_0}}{\ \rho\ }{C_\nh\pt{\bs{x}_0}}\ \times\\
\times\ \prod_{j=1}^{n-1}\bigg[\Big|\bkkk{C_\nh\pt{\bs{x}_j}}{\ U\
}{C_\nh\pt{\bs{x}_{j-1}}}\Big|^2\bigg]\ \mu\pt{\ud \bs{x}_0}\mu\pt{\ud
\bs{x}_1}\cdots\mu\pt{\ud \bs{x}_{n-1}}%
\label{riferimento}
\quad,
\end{multline}
where we have used the \co{normalization} property for the state
$\coh{\nh}{\bs{x}_{n-1}}$ and the notation given in Definition~\ref{stringhe}
for the strings $\bs{i}$.

This quantities can be seen as quantum analogue to the classical
probability $\mu_{\bs{i}}$ of Section~\ref{KSE} (in particular they sum up to one) and thus
can be used in computing a Shannon entropy, depending on the given
dynamics $U$, the instrument~\eqref{inst}, the classical partition
${\c E}$, the initial state $\rho$ and the considered time of measuring
$n$, whose expression is
\begin{equation}
S(U,{\cal
I,E},\rho,n)\coleq-\sum_{\bs{i}\in\Omega_D^n}{\c
P}_{\bs{i}}^{\text{\co{CS}}}
\log{\c P}_{\bs{i}}^{\text{\co{CS}}}\quad.
\label{CSQDE_2}
\end{equation}
The \co{CS} \enfasi{quantum entropy}~\cite{Slo94:1} is defined as the
``average production'' on the long run of last quantity
\begin{equation}
H(U,{\cal I,E},\rho)\coleq \lim_{n\to\infty}\frac{1}{n} \;S(U,{\cal
I,E},\rho,n)%
\label{karolen}
\end{equation}
and it is decomposable in two component. The first, called
\enfasi{measurement} \co{CS} 
\enfasi{quantum entropy}, is independent on the
dynamics, originated by the pure measurement process, and obtained by
replacing the unitary operator $U$ in~\eqref{karolen} with the
identity on ${\c H}$; its expression is
\begin{align}
H_{\text{meas}}({\cal I,E},\rho)&\coleq H(\Id_\nh,{\cal I,E},\rho)\quad\cdot%
\label{e_misura}
\intertext{The second amount to the remaining part}
H_{\text{dyn}}(U,{\cal I,E},\rho)&=H(U,{\cal
I,E},\rho)-H_{\text{meas}}({\cal I,E},\rho)%
\label{e_dinamica}
\end{align}
and is supposed to incorporate the dynamic dependence.
\subsection{\co{CS} Entropies for discrete classical systems}%
\label{CSDE}
The quantum entropy of last section can be seen as an
algebraic quantity, and does need nothing more that the algebraic
framework already developed in Sections~\ref{CDS}--\ref{CCLD}, in
order to be defined. In particular, we are going to estimate the
\co{CS} entropy of discrete classical systems $\tripQS$, using
the \co{L}attice \co{S}tates of Definition~\ref{xxxnnn}\ \\[-5ex]
\begin{quote}
\begin{TT}{:} %
\label{TEOCS}
 Let $\pt{\IT,\mu,T}$ be the classical dynamical system of Section~\ref{CDS},
 which is the continuous limit of a sequence of finite dimensional
 discrete dynamical systems $\tripQS$. If\\[-5ex]
 \begin{enumerate}
 \item $W_{T,N}$ is the unitary evolution operator of~\eqref{aggiunta};
 \item ${\c I}$ in the instrument~\eqref{inst} constructed with the
 \co{LS} of Definition~\ref{xxxnnn};
 \item
  $\c E = \{ E_0, E_1,\ldots, E_{D-1} \}$ is a finite measurable
  partition of $\IT$;
 \item $\rho$ is the tracial state $\frac{1}{\nh} \;\Id_{\nh}$;
\\[-5ex]
 \end{enumerate} 
 then there exists an $\alpha$ such that
 \begin{equation*}
  \lim_{\substack{n,N\to\infty\\
n<{\alpha}\:{\log N}}}  \frac{1}{n}
  \left| S(W_{T,N},{\cal
I,E},\rho,n) - S_\mu(\c E_{[0,n-1]}) \right| = 0\quad\cdot
 \end{equation*}
\end{TT}
\end{quote}
In order to prove Theorem~\ref{TEOCS}, we need the following 
auxiliary result.\\[-3.5ex]
\begin{quote}
\begin{LLL}{}\ \\[-2.5ex]
\label{ultimolemma}
Suppose to have a
sequence $\pg{g_N}$ of $\Lspace{2}{\IT}$ functions
such that \mbox{$\norm{g_N}{2}\leq 1$}, $\forall\;N\in\IN^{+}$
($\norm{\cdot}{2}$ meaning the $\Lspace{2}{\IT}$--norm).
\bigskip

Using the quantities $K_{N,n}\pt{\bs{x},\bs{y}}$ of
Definition~\ref{kappaj} we have that,
for any given $A$ and $B$ measurable subsets of $\IT$, and $N$ large enough,
it holds
\begin{align}
R_N& \coleq\abs{\int_{B}\mu\pt{\ud
\bs{x}}g_N\pt{\bs{x}}\nh\int_{A}\mu\pt{\ud\bs{y}}
{\abs{K_{N,1}\pt{\bs{x},\bs{y}}}}^2-
\int_{B\;\cap\; T^{-1}\pt{A}}\mu\pt{\ud
\bs{x}}g_N\pt{\bs{x}}
}
\nonumber\\
& \leq\varepsilon_{B}\pt{N}\quad,\nonumber
\end{align}
where $\varepsilon_{B}\pt{N}\longrightarrow 0$ with $N\longrightarrow
\infty$\ .
\end{LLL}\ \\[-5.5ex]
\end{quote}\ \\[-10ex]

\noindent The symbol $\varepsilon_B$ does not imply any dependence of the
 bounding term
 $\varepsilon_B$ 
\mbox{on the subset $B$;} it is just a way of writing that will be of use in
the following.

\noindent\textbf{Proof of Lemma~\ref{ultimolemma} :}\\[1ex]
\noindent Resorting to the use of the characteristic functions ${\c
X}_A$ and ${\c X}_B$, using triangular inequality and collecting
terms, $R_N$ can be rewritten as
\begin{align}
R_N & \leq
\int_{\IT}\mu\pt{\ud
\bs{x}}\Big|{\c X}_B \pt{\bs{x}}g_N\pt{\bs{x}}\Big|\cdot\abs{
\nh\int_{\IT}\mu\pt{\ud\bs{y}}{\c X}_A \pt{\bs{y}}
{\abs{K_{N,1}\pt{\bs{x},\bs{y}}}}^2-
{\c X}_{T^{-1}\pt{A}} \pt{\bs{x}}
}
\nonumber\\
& =
\norm{\;{\c X}_B \;g_N\;\pq{
\nh\int_{\IT}\mu\pt{\ud\bs{y}}{\c X}_A \pt{\bs{y}}
{\abs{K_{N,1}\pt{\;\bs{\cdot}\;,\bs{y}}}}^2-
{\c X}_A \pt{T\pt{\bs{\;\cdot\;}}}
}}{1}
\nonumber\quad,
\intertext{and using the Cauchy--Schwartz inequality}
& \leq
{\Big\|\;{\c X}_B \;g_N\;\Big\|}_2\cdot\norm{
\nh\int_{\IT}\mu\pt{\ud\bs{y}}{\c X}_A \pt{\bs{y}}
{\abs{K_{N,1}\pt{\;\bs{\cdot}\;,\bs{y}}}}^2 -
{\c X}_A \pt{T\pt{\bs{\;\cdot\;}}}
}{2}
\label{dimlem_1}
\quad\cdot
\end{align}
Now we use the hypothesis, so that 
\begin{equation}
{\Big\|\;{\c X}_B \;g_N\;\Big\|}_2^2=\int_{B}{\Big|
g_N\pt{\bs{x}}\Big|}^2\;\mu\pt{\ud\bs{y}}\leq
{\Big\|\;g_N\;\Big\|}_2^2\leq 1\quad\cdot
\label{dimlem_2}
\end{equation}
Putting together~\eqref{dimlem_1} and~\eqref{dimlem_2}, and using
Proposition~\ref{deltadirac} (with $f={\c X}_A$ and $n=1$) we get the
result.\hfill$\qed$ \\[1.5ex]   
We are now in position to conclude with:\\[1.5ex]
\noindent\textbf{Proof of Theorem~\ref{TEOCS} :}\\[1.5ex]
Let us start to compute the expectation ${\c P}_{\bs{i}}^{\text{\co{CS}}}$.
In terms of the quantity introduced in points $(1$--$4)$ of the
statement, equation~\eqref{riferimento} can be rewritten as 
\begin{align}
{\c P}_{\bs{i}}^{\text{\co{CS}}}& =
\nh^{n-1}\int_{E_{i_0}}\int_{E_{i_1}}\cdots\int_{E_{i_{n-1}}}
\bkkk{C_\nh\pt{\bs{x}_0}}{\ \Id_{\nh}\ }{C_\nh\pt{\bs{x}_0}}\ \times\nonumber\\
& \times\ \prod_{j=1}^{n-1}\bigg[\Big|\bkkk{C_\nh\pt{\bs{x}_j}}{\ W_{T,N}\
}{C_\nh\pt{\bs{x}_{j-1}}}\Big|^2\bigg]\ \mu\pt{\ud \bs{x}_0}\mu\pt{\ud
\bs{x}_1}\cdots\mu\pt{\ud \bs{x}_{n-1}}\quad\nonumber\\
\intertext{and using \co{normalization} property for the state
$\coh{\nh}{\bs{x}_0}$ and resorting to Definition~\ref{kappaj}}
& =
\int_{E_{i_{n-1}}}\cdots\int_{E_{i_1}}\int_{E_{i_0}}\mu\pt{\ud \bs{x}_{n-1}}
\times\
\prod_{j=1}^{n-1}\bigg[\nh\;\Big|K_{N,1}\pt{\bs{x}_{j},
\bs{x}_{j-1}}\Big|^2\mu\pt{\ud \bs{x}_{j-1}}\bigg]\ %
\label{TEO3_1}
\quad\cdot
\end{align}
Now it start an iterate procedures, consisting of two points.\\[1.5ex]
{$\bs{1)}$}\quad consider the function
\begin{align}
g_N\pt{\bs{x}_1}&\coleq\int_{E_{i_{n-1}}}\cdots\int_{E_{i_3}}\int_{E_{i_2}}
\ \prod_{j=2}^{n-1}\bigg[\nh\;\Big|K_{N,1}\pt{\bs{x}_{j},
\bs{x}_{j-1}}\Big|^2\mu\pt{\ud \bs{x}_{j}}\bigg]\ %
\label{TEO3_2}
\quad:
\intertext{all the factors inside the integrals of~\eqref{TEO3_2} are positive,
so that extending the integration domain and expliciting the form of $K_{N,1}\pt{\bs{x}_{j},
\bs{x}_{j-1}}$, we get the bound}
g_N\pt{\bs{x}_1}&\leq\int_{\IT}\cdots\int_{\IT}\int_{\IT}
\ \prod_{j=2}^{n-1}\bigg[\nh\;\Big|\big\< C_\nh(\bs{x}_{j})\:,\:
W_{T,N}\, C_\nh(\bs{x}_{j-1})\big\>\Big|^2\mu\pt{\ud
\bs{x}_{j}}\bigg]=1
\nonumber
\end{align}
from \co{completeness} and \co{normalization}, so that it
follows \mbox{$\norm{g_N}{2}\leq 1$}.\\[1.5ex]
{$\bs{2)}$}\quad 
By means of~\eqref{TEO3_2}, equation~\eqref{TEO3_1} can be rewritten as
\begin{equation}
{\c P}_{\bs{i}}^{\text{\co{CS}}}=\int_{E_{i_1}}\mu\pt{\ud
\bs{x}_1}g_N\pt{\bs{x}_1}\nh\int_{E_{i_0}}\mu\pt{\ud\bs{x}_0}
{\abs{K_{N,1}\pt{\bs{x}_1,\bs{x}_0}}}^2
\nonumber\quad\cdot
\end{equation}
Now Lemma~\ref{ultimolemma} guarantees that there exists a positive sequence
$\varepsilon_{E_{i_1}}\pt{N}$ such that, 
\begin{equation}
\abs{{\c P}_{\bs{i}}^{\text{\co{CS}}}-
\int_{E_{i_1}\;\cap\; T^{-1}\pt{E_{i_0}}}\mu\pt{\ud
\bs{x}_1}g_N\pt{\bs{x}_1}
}\leq\varepsilon_{E_{i_1}}\pt{N}
\nonumber\quad,
\end{equation}\ \\[3ex]
with $\varepsilon_{E_{i_1}}\pt{N}\longrightarrow 0$ for $N\longrightarrow
\infty$\ .
By iterating $(n-1)$--times this procedure (consisting in isolating a single
$K_{N,1}\pt{\bs{x}_{j},\bs{x}_{j-1}}$ and grouping all the others in a
single bounded function $g_N\pt{\bs{x}_j}$) and using the triangle
inequality for $\abs{\cdot}$, we finally arrive to the result:
\begin{gather}
\abs{{\c P}_{\bs{i}}^{\text{\co{CS}}}-
\mu\pt{E_{i_{n-1}}\cap
T^{-1}\pt{E_{i_{n-2}}}\cap\cdots\cap T^{1-n}\pt{E_{i_{0}}}}
} 
= \abs{{\c P}_{\bs{i}}^{\text{\co{CS}}}-
\mu_{\bs{\hat{\imath}}}^{\phantom{\text{\co{C}}}}
}
\leq \varepsilon\pt{N}
\quad,%
\nonumber%
\intertext{with}
\varepsilon\pt{N}\coleq \sum_{\ell=1}^{n-1}
\varepsilon_{E_{i_{\ell}}}\pt{N}\longrightarrow 0\quad\text{for}\quad N\longrightarrow
\infty\quad,%
\label{TEO3_5}
\end{gather}
$\mu_{\bs{j}}$ meaning the classical probability of Section~\ref{KSE}
and $\bs{\hat{\imath}}$ denoting the string $\bs{i}$ reversed,
as in Definition~\ref{stringhe}.2.

We now define two \enfasi{density matrices}, with the aim to compute
their \enfasi{Von Neumann Entropy} (see Section~\ref{QDE}), that are both diagonal in the basis
${\pg{\ket{\bs{i}}}}_{\bs{i}\in\Omega_D^n}$ of the $D^n$ dimensional
Hilbert space ${\c H}_{D^n}$:
\begin{equation}
\rho\coleq\sum_{\bs{i}\in\Omega_D^n}\mu_{\bs{\hat{\imath}}}\ket{\bs{i}}\bra{\bs{i}}\qquad,\qquad\sigma\coleq\sum_{\bs{i}\in\Omega_D^n}{\c P}_{\bs{i}}^{\text{\co{CS}}}\ket{\bs{i}}\bra{\bs{i}}\quad\cdot\nonumber
\end{equation}
Resorting to the \enfasi{trace norm}
$\norm{A}{1}\coleq\Tr\abs{A}=\Tr\sqrt{A^{\dagger}A}$, we
use~\eqref{TEO3_5} to estimate $\norm{\rho-\sigma}{1}$, that is
\begin{equation}
\Delta\pt{n}\coleq\norm{\rho-\sigma}{1}\leq D^n \varepsilon\pt{N}
\nonumber
\end{equation}
Finally, by the continuity of the von~Neumann entropy~\cite{Fan73:1}, we get 
\begin{equation*}
 \left|H\pt{\rho}-H\pt{\sigma}\right|
\leq \Delta(n)\log D^n\,+\,\eta(\Delta(n))\quad,
\end{equation*}
that is $\left| S(W_{T,N},{\cal
I,E},\rho,n) - S_\mu({\c E_{[0,n-1]}}) \right|\leq \Delta(n)\log
D^n\,+\,\eta(\Delta(n))$,
indeed the two Von Neumann entropy $H\pt{\rho}$ and $H\pt{\sigma}$ are
nothing but the Shannon entropy of the refinements ${\c
E_{[0,n-1]}}$ of the classical partition (see Section~\ref{KSE}),
respectively the Shannon
entropy~\eqref{CSQDE_2} leading to the \co{CS} quantum entropy.

Since, from $n \le \alpha \log N$, $D^{n}\leq N^{\alpha\log D}$, if
we want the bound $D^{n}\varepsilon(N)$  to converge to zero with
$N\longrightarrow\infty$, the parameter $\alpha$ has to be chosen
accordingly.\hfill$\qed$ \\[3ex] 
By means of Theorem~\ref{TEOCS}, a positive \co{CS}--entropy production is then
associated to discrete systems whose continuous limit exhibit a positive
\co{KS}--entropy production, which correspond in turn to the
sum of all positive Lyapunov exponent of the continuous classical
system, as stated by the Pesin's
Theorem~\cite{Man87:1}. 

This positive \co{CS}--entropy production is entirely due to the
dynamical component $H_{\text{dyn}}(W_{T,N},{\cal I,E},\rho)$
of~\eqref{e_dinamica}, being the measurement
\co{CS}--entropy~\eqref{e_misura} equal to zero, as stated in the next proposition:\ \\[-4.5ex]
\begin{quote}
\begin{PPP}{} %
\label{ent_mis}\ \\[-2ex]
 Let ${\c I}$ and $\c E$ be the
 instrument, respectively the finite measurable
  partition of the statement of Theorem~\ref{TEOCS} and let $\rho$ be
 the tracial state $\frac{1}{\nh} \;\Id_{\nh}$. There exists an
 $\alpha^{\prime}$ such that:
 \begin{equation}
  \lim_{\substack{n,N\to\infty\\
n<{\alpha^{\prime}}\:{\log N}}}  \frac{1}{n}\;
  S(\Id_{\nh},{\cal I,E},\rho,n)= 0\quad\cdot
\nonumber%
 \end{equation}
\end{PPP}
\end{quote}\ \\[-7ex]
\noindent\textbf{Proof:}\\[1.5ex]
Performing a proof completely analogous to the one for
 Theorem~\ref{TEOCS}, we find an $\alpha^{\prime}$ such that
 \begin{equation}
  \lim_{\substack{n,N\to\infty\\
n<{\alpha^{\prime}}\:{\log N}}}  \frac{1}{n}
  \left| S(\Id_{\nh},{\cal
I,E},\rho,n) - S_\mu(\c E^{\prime}_{[0,n-1]}) \right| = 0\quad,
\label{ul_pr_1}
 \end{equation}
with ${\c E}^{\prime}_{[0,n-1]}$ now given by ${\cal E}^{\prime}_{[0,n-1]}
 \coleq\bigvee_{j=0}^{n-1}\Id^j({\cal E})={\cal E}\bigvee{\cal
 E}\bigvee\cdots\bigvee{\cal E}$ (see Section~\ref{KSE}), so that 
\begin{equation}
S_\mu(\c
 E^{\prime}_{[0,n-1]})=S_\mu(\c E)\leq\log D \quad,
\label{ul_pr_2}
\end{equation}
independent of $n$.\\
Now we use triangular inequality together with~\eqref{ul_pr_2}, obtaining
\begin{equation}
\frac{1}{n}\;
S(\Id_{\nh},{\cal
I,E},\rho,n)\leq
\frac{1}{n}
  \left| S(\Id_{\nh},{\cal
I,E},\rho,n) - S_\mu(\c E^{\prime}_{[0,n-1]}) \right|
+ \frac{\log D}{n}\quad,
\label{ul_pr_3} 
\end{equation}
and so the result follows
from~\eqref{ul_pr_1}.\hfill$\qed$%
\section{Conclusions}
\label{concl}
In this work we studied the footprints
of chaos present in classical dynamical systems on the two dimensional
torus after a discretization has forced these systems to move on a
regular lattice of spacing $\frac{1}{N}$, with
finite number of sites $N^2$.

Discretizing is similar to quantizing; in particular, as for the
classical limit $\hbar\rightarrow 0$, we have set up a solid
theoretical framework to discuss the continuous limit
$N\rightarrow\infty$. 

Inspired by the
semi--classical analysis, we developed an
algebraic discretization technique by mimicking the well known
\co{A}nti--\co{W}ick schemes of quantization, in particular we made use of a
family of suitably defined \co{L}attice \co{S}tates with properties
that, in a quantum setting, are typical of \co{C}oherent \co{S}tates.

The result is the appearance of a logarithmic
time--scale when the discrete hyperbolic systems tend to their
continuous limit; namely, the continuous and discrete dynamics agree
up to a breaking time which is proportional to the logarithm
of the lattice spacing.

We also used the entropy production as a
parameter of chaotic behaviour. In particular the notion of
\co{CS}--quantum entropy has been used: this reproduce the classical
metric entropy of \co{K}olmogorov and \co{S}inai if applied to
classical continuous systems. 

The \co{CS}--quantum entropy do converge to the \co{KS} invariant, but on
logarithmic time scales too.  
\newpage
\noindent\textbf{\large Acknowledgments}

\noindent The author wishes to thank Dr. F. Benatti for stimulating
discussions and useful advice. 

\appendix
\section{Sketch of the proofs of Propositions~\ref{prop_hyp},~\ref{prop_par} and~\ref{prop_ell}  }
\label{app_000}
\noindent\textbf{Proof of Proposition~\ref{prop_hyp} :}\\[1ex]
{\bf 1)} --- Let us start by considering matrices with positive trace,
that is positive eigenvalues $\pt{\lambda,\lambda^{-1}}$; the case of
negative trace will be considered in next point {\bf (2)}.
In the (non--orthogonal) reference system $(\bs{\hat{c}}_1,\bs{\hat{c}}_2)$ oriented along
eigenvectors $\pt{\ket{\bs{e}_+},\ket{\bs{e}_-}}$, the time--evolution
is described by 
\begin{equation}
\pt{c_1,c_2}\xrightarrow[n\in\IN]{\quad
T^{\pm n}\quad}\pt{\lambda^{\pm n} c_1,\lambda^{\mp n}c_2} \quad, 
\label{evolutioncc}
\end{equation}
thus orbits are simply given by $c_1 c_2 =$Const., that in the reference
system $\pt{\bs{\hat{x}},\bs{\hat{y}}}$ reads as~\eqref{orbits_hyp},
indeed the relation between coordinates in the two systems is:
\begin{equation}
\begin{pmatrix}
x\\
y
\end{pmatrix}=
\begin{pmatrix}
1 & \cos\beta\\
0 & \sin\beta
\end{pmatrix}
\begin{pmatrix}
c_1\\
c_2
\end{pmatrix}\quad\cdot
\label{camb_hyp}
\end{equation}
Among these orbits, we choose the two that are tangent (and so
closest) to the unit ball $B_T(0)$: of course they remain tangent and
closest even during evolution $B_T(0)\longmapsto B_T(n)$ and so they
give us the the right expression for the surrounding orbits of
$B_T^{(n)}$, that is~\eqref{hyp_hyp}.

By means of~\eqref{evolutioncc} and~\eqref{camb_hyp} we have an
expression for the $\pm n$-evolved unit ball, that is
${B}_T(n)$; among its surface's points we choose the farthest ones and
we determine their norm, getting the expression for ${D}_T(n)$
contained in~\eqref{lun_hyp}.

Now we use the expression $\sinh^{-1}\pt{q}=\log\pt{\sqrt{q^2 + 1} +
q}$, that holds for all $q>0$, in particular for
$q=\pt{\lambda^{n}-\lambda^{-n}}/{\sin\beta}$ ($\sin\beta>0$), so that
from~\eqref{lun_hyp} we get for ${D}_T(n)$ the expression given
by~\eqref{lun_hyp2}, that shows the monotonicity in $n$ of this
function; this monotonicity, together with the
definitions~\eqref{pallamnn} of $B_T^{(n)}$, give us the
equivalence between $D_T^{(n)}$ and ${D}_T(n)$.

The linear matrix action $T$ map the unit ball ${B}_T(0)$ in the
ellipse ${B}_T(1)$ an ${D}_T(1)$ is its major semi--axis; from
Definition~\ref{palle}, we have 
\begin{equation*}
\eta^2=\sup_{\ket{\bs{v}}\in\IR^2}\big\<\bs{v}\big\vert T^\dagger
T\big\vert\bs{v}\big\> =
\sup_{\ket{\bs{v}}\in\IR^2}
\Big\|T\big\vert\bs{v}\big\>\Big\|_{\IR^2}^2 = \pq{{D}_T(1)}^2\nonumber\quad,
\end{equation*}
so that $\eta={D}_T(1)$ and~\eqref{eta_beta_hyp} follows from
expression~\eqref{lun_hyp}, with $n=1$.\\
Expressions in~\eqref{asin_hyp} can be easily deduced
from~\eqref{lun_hyp}.

{\bf 2)} --- Let us now notice that every map $T$, whose trace is
negative, may be written as the composition of $-\Id_2$ (the identity
map) with the map $-T$, which has positive trace; the same holds true
for the iterates ${\pg{T^k}}_{k \text{ odd}}$. Since multiplying by
$-\Id_2$ amounts to perform the transformation
$\pt{x,y}\longmapsto\pt{-x,-y}$, both the orbits~\eqref{orbits_hyp} and the
surrounding surface~\eqref{eta_beta_hyp}, which exhibit a central
symmetry, remain the 
same also for negative trace maps. The same argument can be applied
to the diameter ${D}_T(n)$ of~\eqref{lun_hyp}, which are invariant for
coordinates reflection too.\hfill$\qed$\\[1ex]
\noindent\textbf{Proof of Proposition~\ref{prop_par} :}\\[1ex]
Let us consider matrices $T$ with $\Tr T=2$, that is
$t=1$, being the case $t=-1$ equivalent, as it is possible to prove in
the same way of point {\bf (2)} of the proof of
Proposition~\ref{prop_hyp}. In the orthogonal reference system
$(\bs{\hat{x}},\bs{\hat{y}})$ of the statement, the action of $T^n$ is
described by a matrix in Jordan canonical form, that is
\begin{equation}
\begin{pmatrix}
x\\
y
\end{pmatrix}\xrightarrow[T^n]{}
\begin{pmatrix}
x^\prime\\
y^\prime
\end{pmatrix}=
\begin{pmatrix}
1 & nJ^\prime\\
0 & 1
\end{pmatrix}
\begin{pmatrix}
x\\
y
\end{pmatrix}\quad,
\label{camb_par}
\end{equation}
where $J^\prime=t_{12}-t_{21}$,
thus orbits are simply given by $y =$Const.
In order to apply the argument of point {\bf (2)} of proof of
Proposition~\ref{prop_hyp}, when $t=-1$, we endow this class of orbits
with a coordinate reflection symmetry, and this leads to
equation~\eqref{orbits_par}. 

Among these orbits, we choose the one that is tangent (and so
closest) to the unit ball $B_T(0)$: of course it remains tangent an
closest even during evolution $B_T(0)\longmapsto B_T(n)$ and so it
give us the the right expression for the surrounding orbit of
$B_T^{(n)}$, that is~\eqref{str_par}.

By means of~\eqref{camb_par} we have an
expression for the $\pm n$-evolved unit ball, that is
${B}_T(n)$; among its surface's points we choose the farthest ones and
we determine their norm, getting the expression for ${D}_T(n)$
contained in~\eqref{lun_par}, with $J=\abs{J^\prime}$.

Using once more the expression $\sinh^{-1}\pt{q}=\log\pt{\sqrt{q^2 + 1} +
q}$, that holds for all $q>0$, in particular for
$q=nJ$ , 
from~\eqref{lun_par} we get for ${D}_T(n)$ the expression given
by~\eqref{lun_par2}; using monotonicity
we get the equivalence $D_T^{(n)}={D}_T(n)$.

From $\eta={D}_T(1)$ (see proof of
Proposition~\ref{prop_hyp}), equation~\eqref{J_par} can be earned from
expression~\eqref{lun_par}, with $n=1$.

Expressions in~\eqref{asin_par1} and~\eqref{asin_par2} can be easily
deduced and verified
from~\eqref{lun_par}.\hfill$\qed$\\[1ex]
\noindent\textbf{Proof of Proposition~\ref{prop_ell} :}\\[1ex]
The semi--trace $t$ of the matrix $T$ can only assume values in
$\pg{-\frac{1}{2},0,\frac{1}{2}}$, indeed all entries of $T$ are
integer and $\abs{t}<1$. We read from equation~\eqref{rotation} that
$t=\cos{\phi}$ and so we have for $\phi$ the only possible values 
$\pg{\pm\frac{2}{3}\pi,\pm\frac{1}{2}\pi,\pm\frac{1}{3}\pi}$; everyone
of these values make the time--evolution periodic, as it can be deduced from
equation~\eqref{rotation}. All these cases are similar; we now prove
the statement for $t=\frac{1}{2}$.

{$\bs{t=\frac{1}{2}}$} --- We have $\phi=\pm\frac{1}{3}\pi$ and
so we get from equation~\eqref{rotation} that $T^3=-\Id_2$. The period
of evolution is six and the sequence of $T$--power is equivalent to
$\Id_2$, $T$, $-T^{-1}$, $-\Id_2$, $-T$, $T^{-1}$, $\Id_2$ and so on.

By using equation~\eqref{pallan} of Definition~\ref{palle} we see that
the sequence ${\pg{B_T(n)}}_{n\in\IN}$ of $n$--evolved ball is
equivalent to $B_T(0)$, $B_T(1)$, $B_T(-1)$, $B_T(0)$, $B_T(1)$,
$B_T(-1)\ldots$, thus, the sequence of diameter
${\pg{D_T(n)}}_{n\in\IN}$, is given by $D_T(0)$, $D_T(1)$,
$D_T(-1)\ldots$.

As argued in the proof of Proposition~\ref{prop_hyp} (point
$1$), $D_T(1)=\eta$; moreover
$D_T(-1)=\eta$ too. Indeed, as the spectra of $\abs{T}$ consists of
the two eigenvalue $\pt{\eta,\eta^{-1}}$, the same is true for the
spectra of $\abs{T^{-1}}$.

Using the last observation, the sequence of diameter becomes
$0$, $\eta$, $\eta$, $0$, $\eta$, $\eta\ldots$ and so
equations~(\ref{lun_ell_a}--\ref{lun_ell_b}) hold true for the case
$t=\frac{1}{2}$.  

The cases $\bs{t=-\frac{1}{2}}$ and $\bs{t=0}$ can be proved
in a similar way.\hfill$\qed$
\singlespacing   		

\end{document}